\documentclass[aip,jcp,reprint,amsmath,amssymb]{revtex4-1}

\usepackage{graphicx} 
\usepackage{dcolumn}
\usepackage{bm}
\usepackage[mathlines]{lineno}

\usepackage[utf8]{inputenc}
\usepackage[T1]{fontenc}
\usepackage{mathptmx}
\usepackage{etoolbox}
\usepackage{gensymb}
\usepackage{siunitx}
\DeclareSIUnit[]\Molar{M}

\draft 

\makeatletter
\def\@email#1#2{%
 \endgroup
 \patchcmd{\titleblock@produce}
  {\frontmatter@RRAPformat}
  {\frontmatter@RRAPformat{\produce@RRAP{*#1\href{mailto:#2}{#2}}}\frontmatter@RRAPformat}
  {}{}
}%
\makeatother

\begin{document}

\title{Substrate-Directed Wetting Layers in Bicontinuous Particle-Stabilised Emulsions} 

\author{Jesse M. Steenhoff*}
\email[]{j.m.steenhoff@uu.nl}
\affiliation{Van 't Hoff Laboratory for Physical and Colloid Chemistry, Utrecht University, Utrecht, The Netherlands}

\author{Martin F. Haase**}
\email[]{*m.f.haase@uu.nl}
\affiliation{Van 't Hoff Laboratory for Physical and Colloid Chemistry, Utrecht University, Utrecht, The Netherlands}

\date{\today}

\begin{abstract}
Bicontinuous interfacially jammed emulsion gels (bijels) facilitate efficient mass transport across multiple length scales due to their interwoven structure of particle-stabilised liquid channels. This unique morphology imparts considerable potential for applications in separation and catalysis, particularly when fabricated \textit{via} solvent-transfer-induced phase separation (STrIPS). STrIPS enables the continuous, large-scale production of nanostructured bijel films on solid substrates, yet the influence of the substrate properties on the formation dynamics and final morphology remains insufficiently understood. In this study, this relationship is elucidated by preparing STrIPS bijel films on silane-functionalised glass substrates with selectively controlled wettability and analysing the resulting structure with confocal microscopy. The results showed the presence of notable wetting layers at the bijel-substrate interface, whose thicknesses could be tuned through the nanoparticle weight fraction. In line with numerical simulations, increasing the substrate hydrophobicity drove a transition from a laminar, water-rich surface layer to a patch-like, progressively oil-rich structure. These findings provide crucial insight into the structure-directing role of substrates in supported bijel films, which aids their application as functional materials.
\end{abstract}

\maketitle

\section{Introduction}
Bicontinuous materials are ideal templates for highly optimised mass transfer, connecting processes that range from the transport of lithium ions in batteries \cite{Huang2015,Werner2018,Sheng2021} to the supply of nutrients in cell tissue scaffolds \cite{Hori2019,Dudaryeva2025,Banerjee2025}. Comprising an intricately interwoven network of two distinct, continuous phases, their structure combines extensive interphase contact with short diffusion pathways throughout the entire material \cite{Scriven1976,Wiesner2023,Xiang2023,Bai2026}. Moreover, the bicontinuous morphology facilitates unrestricted mass transport through the channel networks while lacking the ``dead spaces'' of similar porous materials \cite{Bai2026,Beunen2026,Groisman2026}. These properties enable fast mass transfer over multiple length scales, both internally between its different phases and across the entire structure itself, making bicontinuous materials highly suitable for applications in energy storage \cite{Guo2016,Han2023,Tang2024}, catalysis \cite{Zielasek2006,Hsueh2012,Li2020,Kwon2025}, and separation \cite{Zhou2007,Pang2020,Khan2022}. 

\begin{figure*}
    \centering
    \includegraphics[width=\linewidth]{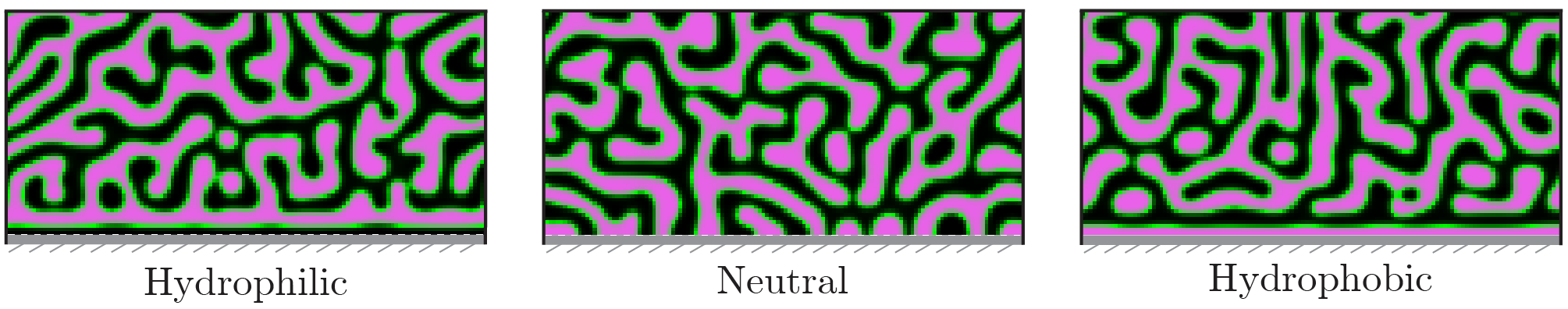}
    \caption{Schematic illustration of the influence of substrate properties on the structure of bijels formed \textit{via} STrIPS. Here, the water- and oil-rich liquid phases of the bijel are indicated in black and magenta, respectively, while the presence of nanoparticles is represented by green. For both very hydrophilic and hydrophobic substrates, surface-directed spinodal decomposition causes the formation of laminar structures at the interface at the bijel-substrate interface, enriched with the water- and oil-rich liquid phase, respectively. In between these laminar extremes there should exist a neutral region, where neither liquid phase shows preferential wetting behaviour.}
    \label{Intro_WettingLayer} 
\end{figure*}
Bicontinuous interfacially jammed emulsion gels (bijels) are a promising subclass of bicontinuous materials \cite{Stratford2005,Herzig2007,Cates2008}, in particular those formed \textit{via} solvent-transfer-induced phase separation (STrIPS) \cite{Haase2015}. During STrIPS, the removal of solvent from a precursor mixture induces the spinodal decomposition of two immiscible liquids. Surfactant-functionalised nanoparticles subsequently attach to the formed liquid-liquid interface, eventually forming a jammed layer that kinetically arrests further phase separation. The result is a bicontinuous network of liquid channels, stabilised by a percolating sheet of nanoparticles. These nanoparticles provide considerable control over the morphology of the bijel, for example through the weight fraction \cite{Tavacoli2011,Witt2013,Steenhoff2026,Steenhoff2026_2}, while simultaneously imparting functionality through their surface chemistry \cite{Haase2017,Vitantonio2019}. Accordingly, STrIPS bijels have found success as template materials for nanocomposite membranes in separation technology \cite{Haase2017,Siegel2022,Siegel2025}, in addition to applications in catalysis \cite{Cha2019,Vitantonio2019} and tissue engineering \cite{Banerjee2025,Okoro2026}.

The STrIPS process is not unique in producing bijels, yet offers a number of distinct advantages over alternatives such as temperature-induced phase separation. Namely, STrIPS bijels can reach smaller domain sizes due to the inclusion of nanoparticles rather than larger colloids \cite{Haase2015,Khan2022}. In addition, they offer greater compositional variety for the liquids in their precursors, not being restricted to highly specific pairs with suitable phase behaviour \cite{Herzig2007,Tavacoli2011,Bai2015,Cai2015}. Most notable, however, is the potential for process up-scaling. Recent work demonstrated that roll-to-roll coating, a technique widely implemented for the industrial production of polymer membranes, can be adapted for the large-scale fabrication of supported STrIPS bijel films \cite{Siegel2024,Siegel2025}. Although this greatly amplifies the potential of STrIPS bijels, it also introduces another factor to their already complex formation dynamics: substrate wetting effects. 

Previously, STrIPS bijels were mainly fabricated with microfluidics; precursor mixtures were injected into a free-standing liquid medium, completing bijel formation before any contact with a solid took place \cite{Sprockel2023,Ruiter2024}. In contrast, the first step of roll-to-roll coating is the deposition of a precursor mixture directly on top of a solid substrate. Consequently, the influence of the substrate on bijel formation has to be taken into account from the start. This was already apparent in earlier work, which established that the production of smooth bijel films requires adequate wetting of the substrate by the precursor mixture \cite{Siegel2024}. Moreover, both the adhesion strength and general structure of the STrIPS bijel films were found to depend on the nature of the substrate material, which was tentatively linked to its influence on the phase separation dynamics \cite{Siegel2025}. However, conclusive evidence for the underlying mechanism remains lacking. 

Fortunately, the associated interplay between wetting and phase separation has been thoroughly explored for other systems, polymer mixtures in particular \cite{Krausch1995,Tanaka2001,Geoghegan2003}. Here, the so-called ``surface-directed spinodal decomposition'' is known to induce the formation of a surface layer enriched with the phase that best wets the substrate \cite{Jones1991,Wang2000,Jinnai2003,Das2020}. The structure of this surface layer can significantly deviate from that of the bulk, even being reported as fully laminar in some cases \cite{Bruder1992,Tanaka1993,Tanaka1993_2,Lin1994,Geoghegan1995,Geoghegan2000,Moffitt2002}. In the context of STrIPS bijels, this concept is visualised in Figure~\ref{Intro_WettingLayer}. For both very hydrophilic and very hydrophobic substrates, the bijel should exhibit a laminar surface layer primarily composed of the liquid phase with the most favourable wetting. For a typical STrIPS precursor, this corresponds to an oil- or water-rich wetting layer, respectively. In addition, there should exist a neutral region in which the wetting layer transitions between the two laminar extremes, with neither liquid phase showing particularly preferential wetting. 

The main goal of this work is to elucidate the influence of solid substrates on the formation of STrIPS bijels by investigating the role of surface layers. Guided by insights from phase-field simulations, this is achieved by fabricating STrIPS bijel films on substrates with varying precursor wettability. To obtain such substrates, glass slides are rendered progressively more hydrophobic through controlled silanisation of their silica surface. The supported bijel films are extensively analysed with laser scanning confocal microscopy, yielding fully-resolved 3D structures. In line with the concept in Figure~\ref{Intro_WettingLayer}, the bijel films exhibit distinct surface layers that match the properties of the underlying substrate. Moreover, these surface layers undergo a gradual compositional transition from predominantly water-rich to oil-rich with increasing hydrophobicity of the substrate. 

\section{Materials \& Methods}
\begin{figure}
    \centering
    \includegraphics[width=\linewidth]{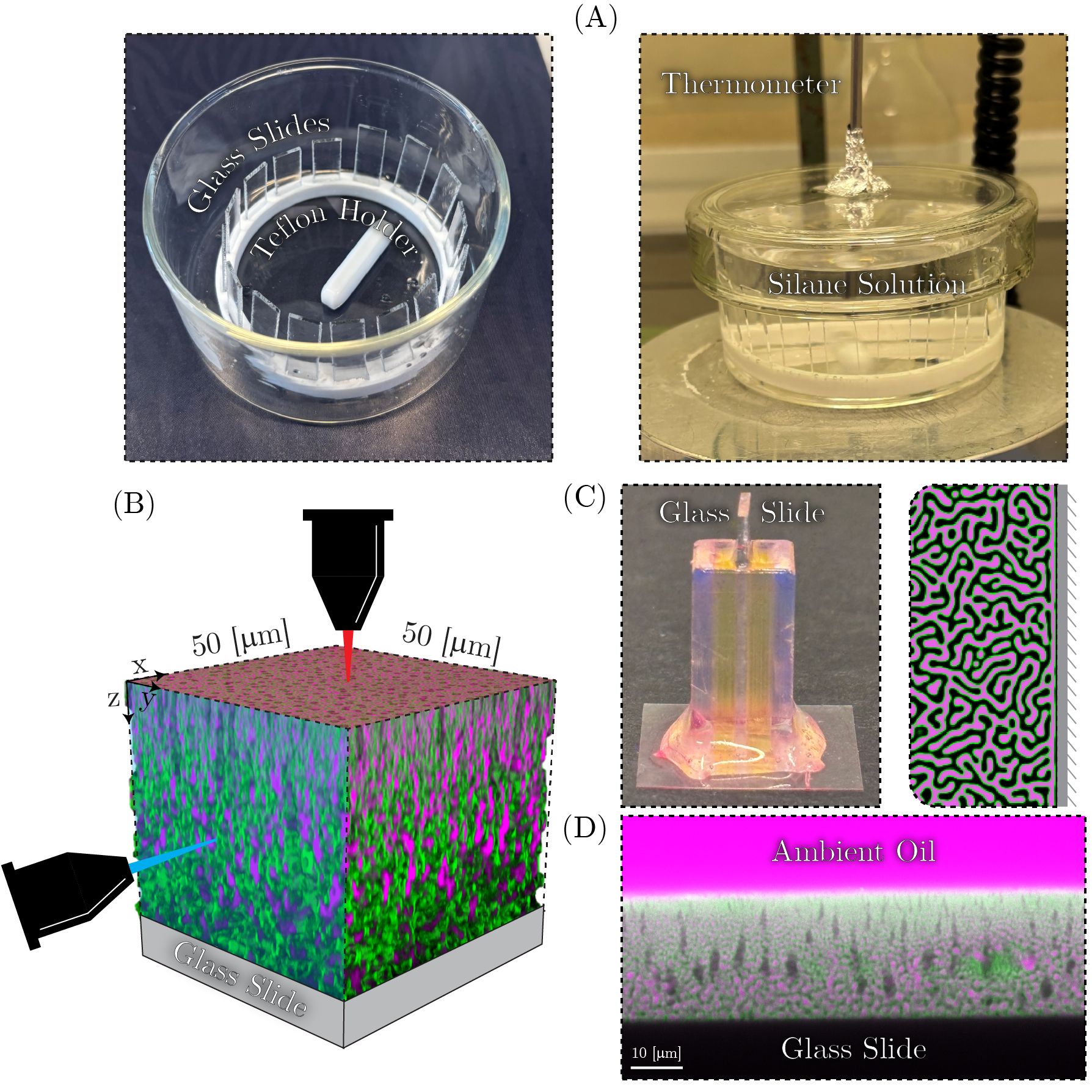}
    \caption{\textbf{(A)} Experimental setup for the controlled silanisation of glass slides, consisting of a custom Teflon ring (left) that allows for stirring and a thermometer (right) for temperature control. \textbf{(B)} 3D confocal reconstructions of a STrIPS bijel film on a glass slide, where the magenta, black, and green colours correspond to oil, water, and nanoparticles, respectively. The red $xy$-plane is imaged during a standard confocal $z$-stack, while the blue $yz$-plane is imaged with a 3D-printed sample container. \textbf{(C)} The 3D-printed sample container for \textit{in situ} imaging of bijel film cross-sections ($yz$-plane), along with a schematic representation of the relative orientations of the glass slide and the bijel film. \textbf{(D)} \textit{In situ} confocal image acquired with the sample container in (C), providing a full vertical cross-section of a supported bijel film in addition to the glass slide and the oil ambient phase.}
    \label{ExperimentalSetup} 
\end{figure}
\subsection{Materials}
All chemicals were used as received. Diethyl phthalate (DEP, \qty{\geq 99}{\percent}), glycerol (\qty{\geq 99}{\percent}), hydrochloric acid (HCl, \qty{37}{w\percent}) and toluene (\qty{\geq 99}{\percent}) were acquired from  Thermo Scientific; Acetic acid (glacial, \qty{100}{\percent}), hexadecyltrimethylammonium bromide (CTAB, \qty{\geq 99}{\percent}), 1-propanol (\qty{\geq 99.5}{\percent}) and Nile red (for microscopy) were purchased from Sigma-Aldrich; Ethanol (Dehydrated, \qty{\geq 100}{\percent}), Methanol (HPLC) and n-hexane (HPLC) were obtained from Biosolve BV; hydrochloric acid (\qty{1}{\Molar}) was acquired from Acros Organics; Silica nanoparticles (Ludox TMA, \#1003481587, particle diameter \qty{22}{\nano\metre}) were received from Grace; Microscope slides (cut, uncoated) were purchased from Epredia. Finally, n-octyltrimethoxysilane (n-OTMS, \qty{ 98}{\percent}) was acquired from Gelest. 

\subsection{Controlled Silanisation of Glass Substrates}
The gradual silanisation of a silica surface with n-OTMS provides a suitable substrate with controlled wettability. To this end, glass microscope slides were cut into smaller \qtyproduct{8 x 25}{\milli\metre} rectangles with a laser-cutter~(GCC X252) and placed in a custom Teflon holder. This slide holder, shown in Figure~\ref{ExperimentalSetup} (A), is a simple Teflon ring with rectangular slots that match the dimensions of the cut glass slides. In contrast to the commonly used Coplin jar, the slide holder allows for continuous stirring during the silanisation process, which minimises gradients in both silane concentration and temperature. 

Prior to silanisation, the glass slides were submerged in a $1:1$ volume ratio of fuming hydrochloric acid and methanol, removing any organic contaminants still present \cite{Cras1999}. Afterwards, the slides were thoroughly rinsed with ultrapure water (Milli-Q) and dried under nitrogen flow. The silane solution is prepared by dissolving \qty{2}{w\percent} n-OTMS in ethanol along with \qty{5}{w\percent} water and \qty{10}{w\percent} acetic acid. Here, the latter two simply serve to speed up the silanisation reaction. The silane solution was then equilibrated at a temperature of \qty{70}{\celsius}, after which the silanisation process was initiated by full immersion of the cleaned glass slides. As shown in Figure~\ref{ExperimentalSetup} (A), the container with the silane solution was then closed from the top to slow the evaporation of ethanol.

To obtain substrates of varying wettability, glass slides were removed from the hot silane solution at times that varied between \qty{5}{\minute} and \qty{3}{\hour} after initial immersion. After rinsing the glass slides with ultrapure water and drying them under nitrogen flow, the achieved hydrophobicity was assessed with optical contact angle (OCA, Dataphysics 25) measurements using \qty{3}{\uL} water droplets. Additionally, the surface chemistry of the treated glass slides was characterised through Fourier transform infrared (FTIR) spectroscopy (PerkinElmer Frontier). The Results section provides a detailed discussion of the relationship found between the extent of glass surface functionalisation and immersion time in the silane solution. 

\subsection{Preparation of Bijel Precursor Mixture}
The composition of the STrIPS bijel precursor is identical to that used in Chapter 4 of this thesis. A detailed description of its preparation, in addition to the exact relationship between the concentration of the surfactant CTAB and the weight fraction of the nanoparticles, can be found there. 

In summary, the preparation of the bijel precursor started with the concentration of an aqueous dispersion of Ludox TMA silica nanoparticles to \qty{50}{w\percent} and lowering its pH to $1.8$ by adding \qty{1}{\Molar} HCl. Immediately after acidification, the nanoparticle dispersion was mixed with a combination of DEP, glycerol, 1-propanol and CTAB to obtain the following composition: $w_{DEP}=0.06$, $w_{1prop}=0.26$, $w_{wat}=0.29$, $w_{gly}=0.09$, $w_{NP}=0.29$, with $w$ as the weight fraction of the component. The weight fraction of the nanoparticles in the final precursor is determined by the concentration of the surfactant CTAB. Namely, by varying the CTAB concentration between \qty{15}{\milli\Molar} and \qty{40}{\milli\Molar}, different degrees of nanoparticle aggregation were induced in the initial mixture. By then sedimenting the nanoparticle aggregates through \qty{30}{\minute} of centrifugation (Microfuge 16, Beckman Coulter, \qty{16160}{\times g}) and isolating the supernatant, stable precursor mixtures were obtained with nanoparticle loadings ranging from \qty{18}{w\percent} to \qty{29}{w\percent}. 
 
\subsection{Preparation \& Confocal Analysis of Supported Films}
In this work, supported bijel films were fabricated by dip-coating functionalised glass substrates. Here, the glass slides were sequentially immersed in bijel precursor and bulk toluene. During the latter immersion, the solvent diffuses from the precursor into the ambient toluene, initiating the formation of a bijel film \textit{via} STrIPS. Afterwards, the samples were made ready for confocal analysis by replacing the toluene with dye-saturated hexane (nile red). 

As illustrated in Figure~\ref{ExperimentalSetup} (B), the confocal analysis of the supported bijel films consisted of imaging two perpendicular planes. The first, indicated as the $xy$-plane, lies parallel to the surface of the glass substrate. This is the standard plane for a so-called confocal $z$-stack, and provides an optical section of the material at a certain depth $z$. By combining images taken at different depths, the full 3D structure of the bijel film can be effectively reconstructed. However, this method suffers from deteriorating signal quality deeper in the bijel film, caused by slight differences in the refractive index between the liquid phases and the nanoparticle scaffold. The loss of signal can make it difficult to fully resolve the 3D morphology of any particular sample while also unequivocally identifying the region of the bijel-substrate interface. 

Here, this issue was addressed by complementary confocal imaging of the $yz-$plane that lies perpendicular to the supporting substrate. The coated glass slides were placed in the custom 3D-printed sample containers depicted in Figure~\ref{ExperimentalSetup} (C), effectively positioning them perpendicular to the normal imaging plane of the confocal microscope. With this set-up, confocal cross-sections can be taken across the full depth of the coated bijel film at relatively constant signal intensity. Figure~\ref{ExperimentalSetup} (D) shows such a cross-section, capturing an overview of the ambient oil, bijel film, and glass support in a single image. Consequently, the described confocal technique enables complementary verification of the results visible in the more common $z-$stacks, in addition to easy identification of the bijel-substrate interface and associated structural features. 

\section{Results}
\subsection{Building Intuition With Phase-Field Models}
Phase-field models are powerful tools for elucidating the dynamics of phase-separating systems. With a proven track-record for resolving both bijel formation \cite{Steenhoff2025,Steenhoff2026} and phase separation under surface fields \cite{Puri1997,Tanaka2001,Das2020,Wang2021,Nestler2022}, phase-field models are ideal for building preliminary intuition about the influence of substrate properties on the emergence and composition of wetting layers in bijel films.   
\subsubsection{Setting Up the Phase-Field Model}
To start, consider a simple binary mixture of immiscible liquids with composition $\phi$, in contact with a solid substrate located at $y=0$. With $\mathcal{F_\phi}(\phi)$ and $\mathcal{F}_I(\phi,\nabla\phi)$ as the bulk and interfacial contributions to the free energy density, respectively, the total Helmholtz free energy $F$ is given by
\begin{equation}
    F=\int_{V}dV \left\{\mathcal{F_\phi}(\phi)+\mathcal{F}_I(\phi,\nabla\phi)\right\},
    \label{FreeEnergyFunctionals}
\end{equation}
where $V$ is the volume of the system. Bulk phase separation is achieved by employing a Flory-Huggins expression $\mathcal{F_\phi}(\phi)$, such that 
\begin{equation}
    \mathcal{F_\phi}(\phi)=f\left(\phi\ln\phi+(1-\phi)\ln{(1-\phi}+\chi\phi(1-\phi)\right),
\end{equation}
where $f=k_BT/V$, in which $k_B$ is the Boltzmann constant and $T$ the absolute temperature, is the characteristic scale of the free energy in the system. In addition, $\chi$ is the interaction parameter between the two liquids, with phase separation occurring for $\chi>2$. Since the liquids considered here are highly immiscible, a value of $\chi=3$ is a suitable choice \cite{Steenhoff2025,Steenhoff2026}. 

The interfacial term $\mathcal{F}_I(\phi,\nabla\phi)$ accounts for the formation of liquid-liquid interfaces, as well as the presence of the solid substrate. Consequently, it consists of two distinct contributions 
\begin{equation}
    \mathcal{F}_I(\phi,\nabla\phi)=\frac{\kappa}{2}\lvert\nabla\phi\rvert^2 +\delta(y)\Phi(\phi).
\end{equation}
Here, the first term arises from the classical Cahn-Hilliard theory, penalising the formation of liquid interfaces $\lvert\nabla\phi\rvert^2$ through the gradient energy parameter $\kappa$. The second term is associated with the presence of the solid substrate, which imposes a short-range surface potential $\Phi(\phi)$ at its location of $y=0$ through the Dirac delta $\delta(y)$. 

The surface properties of the substrate are encoded in the potential $\Phi(\phi)$. One of the simplest choices for such a potential is  
\begin{equation}
    \Phi(\phi)=f\left(a(2\phi-1)+\frac{b}{2}(2\phi-1)^2\right),
\end{equation}
where the parameters $a$ and $b$ then govern the interactions between the substrate and the immiscible liquids \cite{Puri1997,Binder1998,Tanaka2001}. In particular, the sign of $a$ determines the preferential wetting behaviour: the substrate is hydrophilic for $a>0$ and hydrophobic for $a<0$. The parameter $b$, which controls the shape of the potential profile, is of less relevance here and is always kept at $b=0.50$ for the results shown in this work. 
 
With the free energy $F$ from Eq.~(\ref{FreeEnergyFunctionals}), the chemical potential follows from its functional derivative as 
\begin{equation}
    \mu=\frac{\partial \mathcal{F}_\phi}{\partial\phi}+\delta(y)\frac{\partial \Phi}{\partial\phi}-\kappa\nabla^2\phi,
\end{equation}
where the partial derivatives of the bulk-free energy density $\mathcal{F}_\phi$ and the surface potential $\Phi$ are given by 
\begin{equation}
    \frac{\partial \mathcal{F}_\phi}{\partial\phi}=f\left(\ln{\frac{\phi}{1-\phi}}+\chi(1-2\phi)\right);
    \label{Partial1}
\end{equation}
\begin{equation}
    \frac{\partial \Phi}{\partial\phi}=2f\left(a-b(2\phi-1)\right)
    \label{Partial2}
\end{equation}

Considering a system where mass transport is dominated by diffusion — a known oversimplification but sufficient for the purpose of building intuition — the dynamics of the immiscible liquids are described by the Cahn-Hilliard equation 
\begin{equation}
    \frac{\partial\phi}{\partial t}=\nabla\cdot(M\nabla\mu).
    \label{Cahn-Hilliard}
\end{equation}
Under the additional assumption of constant mobility $M$, Eq.~(\ref{Cahn-Hilliard}) further reduces to 
\begin{equation}
    \frac{\partial\phi}{\partial t}=M\nabla^2\left(\frac{\partial \mathcal{F}_\phi}{\partial\phi}+\delta(y)\frac{\partial \Phi}{\partial\phi}-\kappa\nabla^2\phi\right)
    \label{DynamicEquation}
\end{equation}

To solve Eq.~\ref{DynamicEquation} numerically, it is first rendered dimensionless through the following variables 
\begin{equation}
    \tilde{\textbf{x}}=\frac{\textbf{x}}{\lambda}=\textbf{x}\sqrt{\frac{f}{\kappa}};
\end{equation}
\begin{equation}
    \tilde{t}=\frac{t}{\tau}=t\frac{f^2M}{\kappa},
\end{equation}
which employ the naturally emerging spatiotemporal scales of the system: the interfacial width $\lambda=\sqrt{\kappa/f}$ and the corresponding diffusion time $\tau=\kappa/(f^2M)$. In dimensionless form, Eq.~(\ref{DynamicEquation}) then becomes 
\begin{equation}
    \frac{\partial\phi}{\partial \tilde{t}}=\tilde{\nabla}^2\left(\frac{\partial \tilde{\mathcal{F}}_\phi}{\partial\phi}+\delta(\tilde{y})\frac{\partial \tilde{\Phi}}{\partial\phi}-\tilde{\nabla}^2\phi\right),
    \label{DynamicEquation_Dimensionless}
\end{equation}
where the partial derivatives of $\tilde{\mathcal{F}}_\phi=\mathcal{F}_\phi/f$ and $\tilde{\Phi}=\Phi/f$ can be readily determined from Eqs.~(\ref{Partial1}) and (\ref{Partial2}), respectively. 

Subsequently, Eq.~(\ref{DynamicEquation_Dimensionless}) is solved numerically on a $64\times64$ regular grid using a simple finite-difference approach. The Laplacian is discretised with a central difference scheme, and time integration is performed using the forward Euler method. Periodic boundary conditions are applied in the $x$-direction, while no-flux boundary conditions are enforced in the $y$-direction. Numerical stability is ensured by choosing $\Delta\tilde{x}=\Delta\tilde{y}=1$ for the grid spacing and $\Delta\tilde{t}=10^{-4}$ for the time step. 

\subsubsection{Simulation Results}
\begin{figure*}
    \centering
    \includegraphics[width=\linewidth]{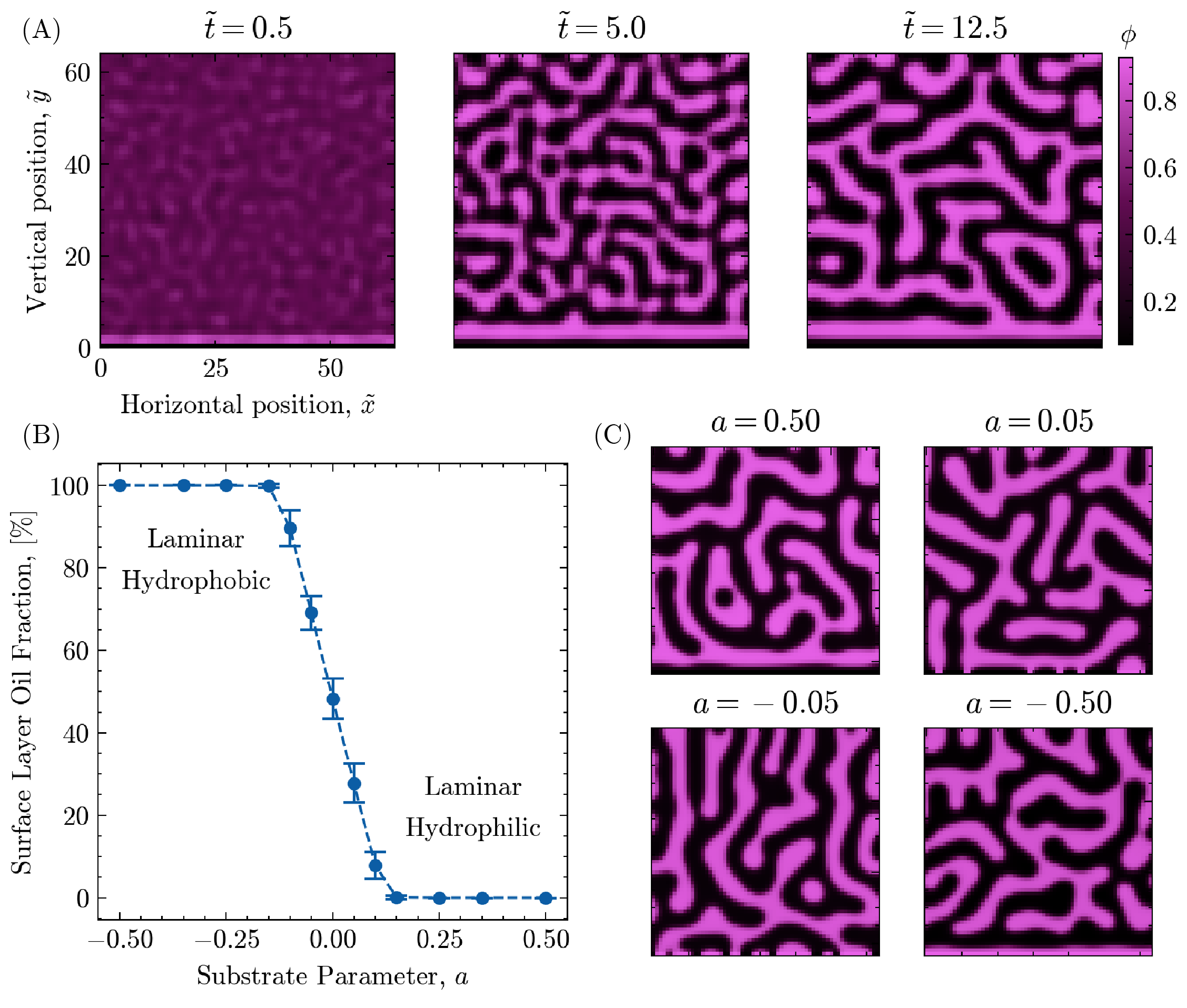}
    \caption{\textbf{(A)} Phase-field simulation showing the spinodal decomposition of two immiscible liquids ($\phi_0=0.50$) in the presence of a hydrophilic substrate located at $\tilde{y}=0$. With a surface parameter $a=0.15$ in Eq.~(\ref{Partial2}) the substrate is strongly hydrophilic, resulting in the formation of a distinctly laminar surface layer consisting entirely of the water-rich phase. \textbf{(B)} Plot showing the fraction of the oil-rich phase in the surface layer at $\tilde{y}=0$ for different values of the substrate parameter $a$. By rendering the substrate increasingly hydrophobic, corresponding to a reduction in $a$, a wetting transition is induced where the composition of the surface layers is inverted from fully water-rich at $a=0.50$ to fully oil-rich at $a=-0.50$. In between, there is an intermediate region where both phases are in contact with the solid substrate. \textbf{(C)} Morphologies of the phase-separated liquid taken at $\tilde{t}=25$ for different values of the substrate parameter $a$. For highly functionalised substrates with $\lvert a\rvert>0.15$, the surface layer is a fully laminar region composed entirely of the preferred liquid phase, the oil-rich and water-rich phase for $a<0$ and $a>0$, respectively. For more neutral substrates with $\lvert a\rvert<0.15$, the surface layer mostly comprises sessile droplets of the preferred liquid phase, partially surrounded by the other. 
}
    \label{PFSimulations} 
\end{figure*}
The simulation results in Figure~\ref{PFSimulations} (A) demonstrate how the presence of a solid substrate can direct the phase separation dynamics of two immiscible liquids. With a symmetric initial composition $\phi_0=0.50$, bulk phase separation occurs through isotropic spinodal decomposition, creating the characteristic structure of labyrinthine channels that is typically associated with bijels. In contrast, the preferential wettability of the substrate is reflected in the strongly anisotropic phase separation directly adjacent to its surface, forming distinctly laminar layers. The substrate here is considerably hydrophilic, $a=0.50$ in the surface potential $\tilde{\Phi}$, resulting in a purely water-rich surface layer. The depletion of water from the bulk forms an oil-rich region directly beyond the surface layer, which is also notably laminar. The layers then coarsen over time, growing in thickness while retaining their laminar structure.

The sequential induction of laminar regions extends the structure-directing influence of the substrate deeper into the bulk than expected based on the short-range interaction imposed by the surface potential $\delta(\tilde{y})\tilde{\Phi}$. Although these numerical simulations do not directly provide a physically meaningful value of this correlation depth, since it depends on a range of factors such as thermal noise and gradient energy parameter, experiments have already shown it can reach up to tens of micrometers \cite{Jinnai2003,Moffitt2002}. Consequently, investigating the formation of surface layers in bijels should be feasible with confocal microscopy.  
 
The relationship between the composition of the surface layer and the properties of the substrate is illustrated in Figure~\ref{PFSimulations} (B). In particular, it shows the fraction of the oil-rich phase in the surface layer as a function of the substrate parameter $a$. As already indicated by the simulations in Figure~\ref{PFSimulations} (A), for hydrophilic substrates with $a>0.15$ the surface layer consists fully of the water-rich phase. By making the surface increasingly hydrophobic, through lower values of $a$, more oil is introduced into the surface layer until there are roughly equivalent amounts of oil and water present at $a=0$. Decreasing the value of $a$ even further effectively renders the substrate hydrophobic, eventually achieving the complete inversion of the surface layer composition for $a<-0.15$. 

In addition to the composition, the structure of the surface layer also varies depending on the nature of the substrate. The combined insights from the plot in Figure~\ref{PFSimulations} (B) and the morphologies in Figure~\ref{PFSimulations} (C) indicate that for both very hydrophilic and very hydrophobic substrates, where $\lvert a\rvert>0.15$, the surface layer is a laminar structure composed fully of the favoured liquid phase. In between these two extremes of surface functionalisation lies an intermediate region where both phases are in contact with the substrate. The morphologies in Figure~\ref{PFSimulations} (C) suggest that in this region the surface layer is formed by droplets of the preferred liquid phase, which are encapsulated by the other phase. As such, this process resembles a partial to complete wetting transition for increasing substrate functionalisation. 

In summary, the results from the phase-field simulations point towards the formation of distinct wetting layers in the presence of a solid substrate. A wetting transition can be induced by varying the surface properties of the substrate, which is reflected in both the composition and general structure of the surface layer. For the experimental system of STrIPS bijel films supported by substrates of varying wettability, confocal microscopy can verify the existence of surface layers and resolve the associated wetting transition. 

\subsection{Surface Functionalisation of Glass Substrates}
\begin{figure*}
    \centering
    \includegraphics[width=\linewidth]{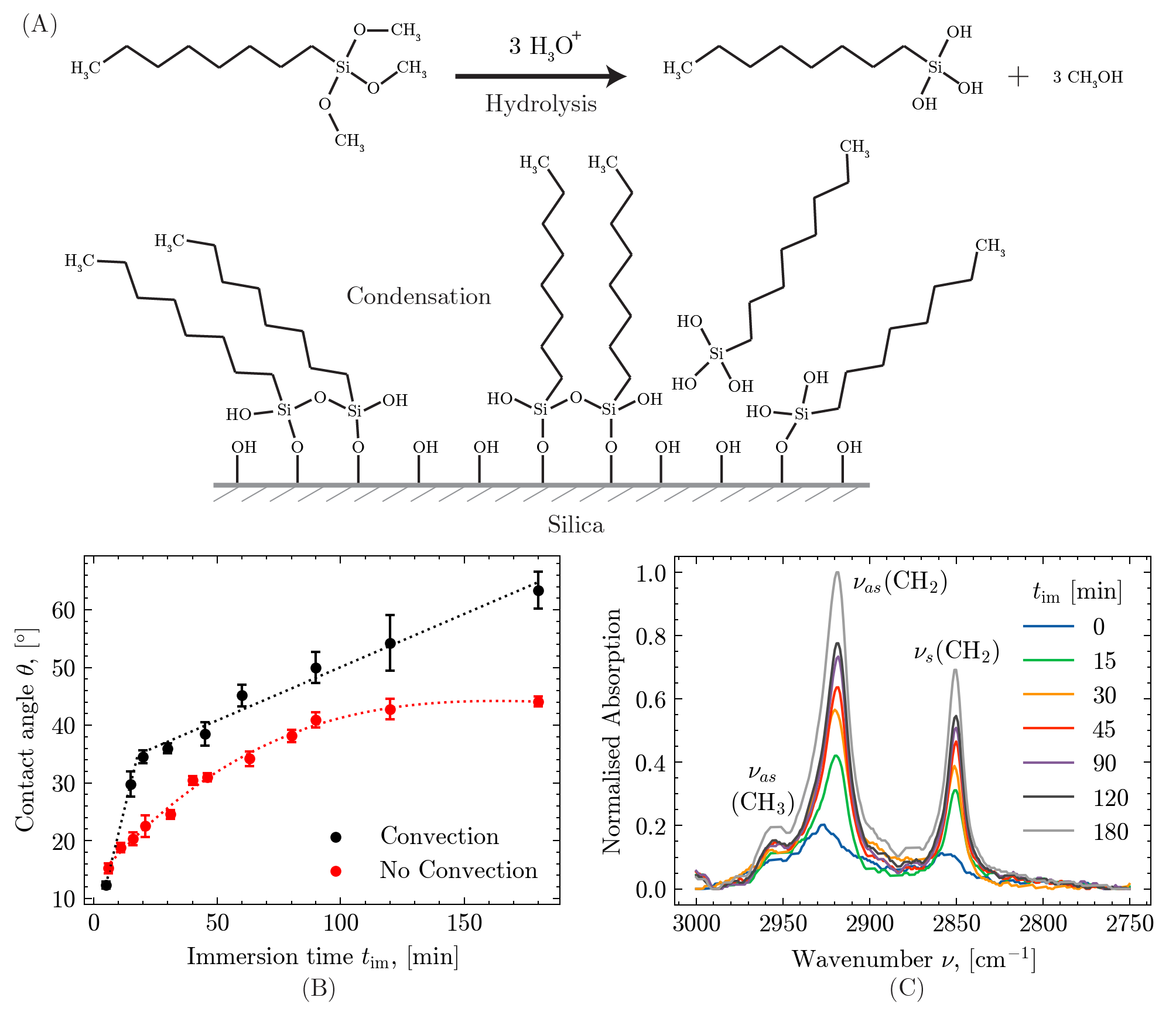}
    \caption{\textbf{(A)} Schematic mechanism of the silanisation of a silica surface using the alkyltrialkoxysilane n-OTMS. First, the three methoxy groups are hydrolysed to silanols in the presence of an acid catalyst. These silanols then undergo condensation reactions with those from other hydrolysed silanes, in addition to the terminal silanols on the silica surface. The former causes the formation of silane oligomers, while the latter covalently grafts the silane to the silica surface. \textbf{(B)} The contact angles of sessile water droplets on functionalised glass substrates against the immersion time in silane solution. The silane solution was continuously stirred for the substrates in the black dataset, while it was not for those in the red dataset. The dotted lines are fitted polynomials meant to guide the eye. \textbf{(C)} FT-IR spectra of the functionalised glass surfaces in the \qtyrange{2750}{3000}{\per\cm} region, taken after different immersion times in the silane solution. The sharp peaks at \qty{2850}{\per\cm} and \qty{2920}{\per\cm} are assigned to the characteristic vibrational modes of the $-\text{CH}_2-$ groups from covalently bound alkane chains, while the broader peak at \qty{2950}{\per\cm} is assigned to the terminal $-\text{CH}_3$ groups.}
    \label{SubstrateSilanisation} 
\end{figure*}
To experimentally validate the results from phase-field simulations, the wettability of glass slides was modified with a silane. A silica surface is highly suitable for silane treatment \cite{Arkles1977}, making glass slides ideal substrates to investigate the influence of wettability on the structure of supported bijel films. Figure~\ref{SubstrateSilanisation} (A) schematically illustrates the mechanism underlying silanisation of a silica surface. In this work, surface functionalisation is achieved through the alkyltrialkoxysilane n-OTMS, which has a long octyl chain in addition to three methoxy groups as its substituents. In solution, the three methoxy groups are first hydrolysed to silanols in the presence of an acid catalyst. The silanol-substituted silane then proceeds to form oligomers and bond with the terminal silanol groups of the silica surface. Eventually, the silane and surface form a siloxane bond through the condensation of their silanol groups, covalently grafting the octyl chain to the surface of the glass slides. 

The substitution of terminal silanols with octyl groups renders the silica surface increasingly hydrophobic. The degree of substitution can be tuned through the immersion time of the glass slides in the silane solution, providing a straightforward method to control their wettability. The effectiveness of this method is demonstrated by the results in Figure~\ref{SubstrateSilanisation} (B), showing the contact angle of sessile water droplets versus the immersion time of the glass substrate. The two colours reflect different experimental conditions: for the glass slides in the black dataset, the silane solution was stirred continuously during treatment, while there was no stirring for those in the red dataset. 

Irrespective of stirring, the contact angles of the sessile water droplets increase with immersion time of the glass substrates. Here, higher contact angles reflect more hydrophobic surfaces. Consequently, the glass becomes increasingly functionalised with silane at longer immersion times. The progression of this surface functionalisation appears to be highly controlled. The contact angle gradually increases from near-zero for the bare glass slide, too low to be measured accurately and therefore omitted from Figure~\ref{SubstrateSilanisation} (B), to a maximum of roughly \qty{63}{\degree} over the course of \qty{3}{\hour}.

Interestingly, stirring the silane solution has a noticeable effect on the surface modification of the glass slides. For a given immersion time, the absence of stirring consistently results in both a lower absolute value and a lower rate of increase for the contact angle. Moreover, this rate decreases over time, with the contact angle reaching a plateau after \qty{3}{\hour} of substrate immersion. In contrast, stirring the silane solution results in a rapid initial increase in the contact angle in the first \qty{20}{\minute}, followed by a slower but sustained increase for the remainder of the immersion time. This discrepancy suggests that mass transport can be a limiting factor in the efficiency of the silanisation process, perhaps the supply of the silane or its bulkier oligomers to the silica surface. 

It is important to note that neither condition achieves complete silane coverage of the silica surface. That is, the contact angle should tend towards a plateau while approaching surface saturation with silane \cite{Flinn1994,Lowe2011,Hasan2016}. However, the upper curve in Figure~\ref{SubstrateSilanisation} (B) shows no signs of doing so, indicating that the silanisation process is still ongoing even after \qty{3}{\hour} of immersion in a stirred solution. Although such a plateau is observed without stirring, the associated value of \qty{44}{\degree} is far too low to correspond to the complete alkane monolayer \cite{Hasan2016}. Rather, the constant contact angle in the absence of stirring could imply the formation of a quasi-steady state with only partial silane coverage of the silica surface. 

Finally, the surface chemistry of the functionalised glass slides was characterised using FT-IR spectroscopy. The results are illustrated in Figure~\ref{SubstrateSilanisation} (C), showing the spectra of the glass surfaces after different immersion times. To facilitate visualisation and comparison, only the absorption in the \qtyrange{2750}{3000}{\per\cm} region is shown here. The spectra in Figure~\ref{SubstrateSilanisation} (C) confirm the successful silanisation of the silica surface, containing distinct peaks associated with the presence of alkane chains. In particular, the peaks at \qty{2850}{\per\cm} and \qty{2920}{\per\cm} correspond to the symmetric and asymmetric vibrational modes of the $-\text{CH}_2-$ groups, respectively, while the weaker peak at \qty{2950}{\per\cm} reflects the fewer terminal $-\text{CH}_3$ groups. The relative intensity of these characteristic peaks increases with longer immersion times, qualitatively indicating a higher grafting density of the alkane chains on the silica surface. Since higher densities of terminal alkane chains make for a more hydrophobic surface, these observations are perfectly in line with the results in Figure~\ref{SubstrateSilanisation} (C). 

\subsection{Imaging Wetting Layers in STrIPS Bijel Films}
\begin{figure*}
    \centering
    \includegraphics[width=\linewidth]{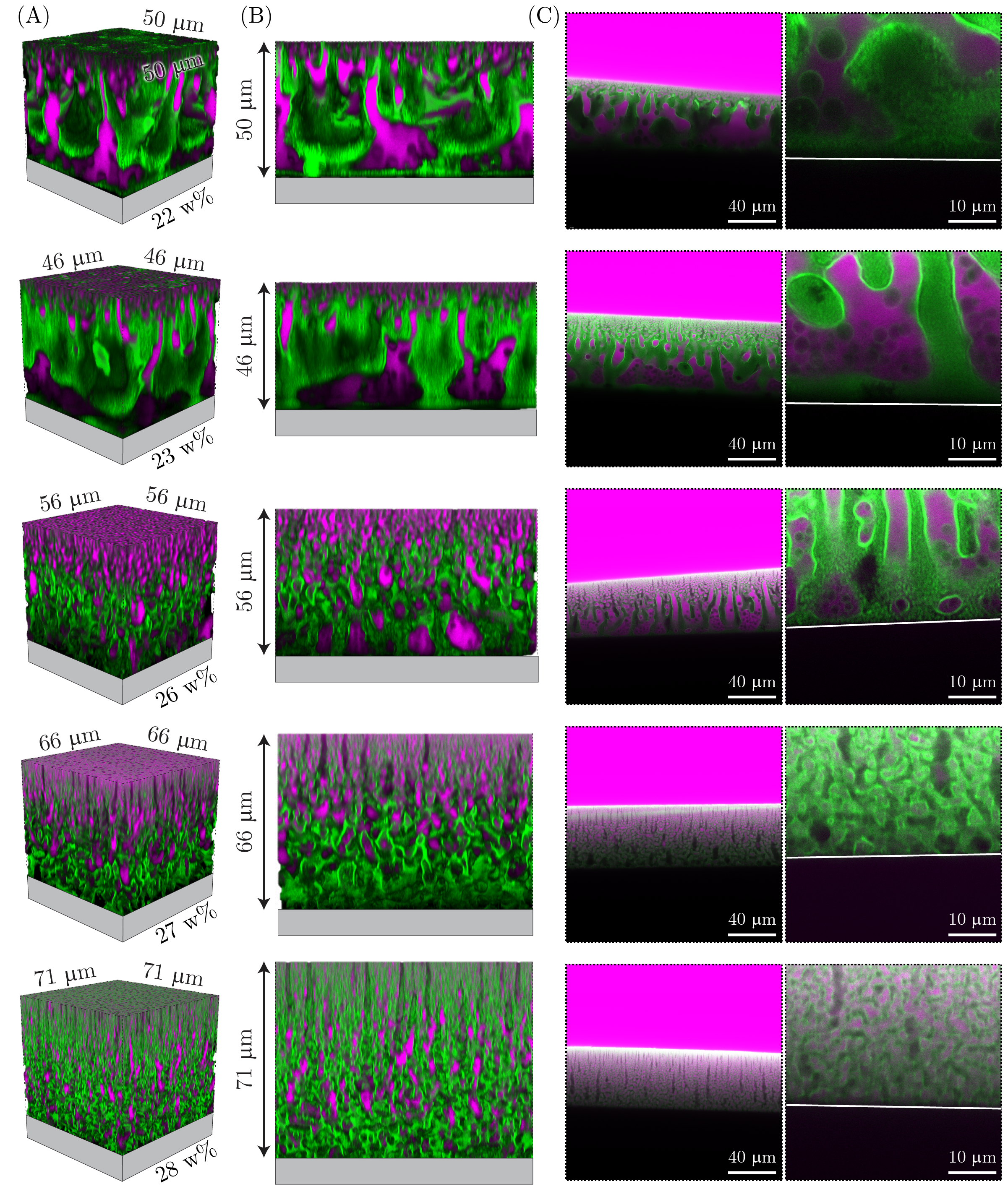}
    \caption{\textbf{(A)} 3D morphologies of supported bijel films with varying nanoparticle weight fractions. The weight fractions of the nanoparticles are provided as percentages. In the images, the magenta signal indicates the presence of oil, while the water is shown in black, and the nanoparticles in green. The supporting substrates are strongly hydrophilic glass slides, exhibiting complete wetting by deposited water droplets. \textbf{(B)} Vertical cross-sections of the bijel films in (A). \textbf{(C)} Complementary confocal images that capture the full vertical cross-sections of supported bijel films in a single image. The right image zooms in on the bijel-substrate interface, which is indicated by the white line. Note that in addition to the water phase, the glass substrate also appears black in these images.}
    \label{NanoparticleLoading} 
\end{figure*}
With precise control over the wettability of the substrate established, the next step was to determine the ideal experimental conditions for the confocal imaging of wetting layers in supported bijels films. In contrast to previously studied systems, the thickness of the wetting layer does not depend solely on the inherent properties of the liquids and substrate. For bijels, the influence of their constituent nanoparticles has to be considered as well. In particular, a larger number of nanoparticles can stabilise a higher interfacial area, so that bijels with higher nanoparticle loading exhibit smaller liquid domains \cite{Herzig2007,Tavacoli2011,Haase2015}. The surface layer, despite its different structure, is unlikely to be an exception. The dynamics of the STrIPS process further complicates the situation, as the resulting bijel films intrinsically exhibit a gradient in the domain size that can be exacerbated by the nanoparticle loading \cite{Steenhoff2025,Steenhoff2026_2}. Consequently, elucidating the relationship between the nanoparticle concentration and the thickness of the wetting layer, and therefore its visibility under the confocal microscope, is of vital importance. 

To achieve this, bijel precursors with different nanoparticle weight fractions were dip-coated on glass slides that were cleaned but not yet silanised. These substrates exhibited complete wetting by deposited water droplets, thus representing the limiting case of a very hydrophilic surface. As such, they provided ideal conditions for observing the formation of a clear, unambiguous wetting layer. The supported bijel films were then analysed by confocal microscopy, using the dual-plane imaging approach described in detail in the Methods section. 

The results of this analysis are shown in Figure~\ref{NanoparticleLoading}. Here, the bijel components are differentiated by colour, with the oil shown in magenta, the water in black, and the nanoparticles in green. Figure~\ref{NanoparticleLoading} (A) depicts the 3D structures of bijel films with different nanoparticle loadings, which were reconstructed from the horizontal cross-sections in confocal $z$-stacks. These structures display all characteristic features of the STrIPS bijel, consisting of an interwoven network of liquid channels with a pronounced gradient in the domain size across the depth of the film. In addition, they illustrate the general influence of the nanoparticle content on the morphology of the bijel film: higher nanoparticle weight fractions yield smaller liquid domains. 

To better visualise the bijel-substrate interface, Figure~\ref{NanoparticleLoading} (B) shows vertical cross-sections of the bijel films in Figure~\ref{NanoparticleLoading} (A). However, 3D reconstructions from confocal $z$-stacks suffer from deteriorating signal quality deeper in the bijel structure, particularly for the oil phase. To address this, Figure~\ref{NanoparticleLoading} (C) provides complementary cross-sections acquired with the \textit{in situ} imaging method, capturing the full vertical structure of the bijel film in a single view. 

Together, these two sets of images confirm the presence of a distinct wetting layer for the supported bijel film. Consistent with the strongly hydrophilic nature of the glass surface, the wetting layer appears to be composed entirely of the water-rich phase. This is particularly noticeable at lower weight fractions of the nanoparticles, showing a green/black laminar region directly adjacent to the substrate. The green signal in the black layer originates from the nanoparticles residing in the water-rich phase. Since the glass substrate also appears black in the cross-sections of Figure~\ref{NanoparticleLoading} (C), the green signal from the nanoparticles actually facilitates identifying the composition of the surface layer. 

Finally, the thickness of the wetting layer decreases with increasing nanoparticle loading, in agreement with the general trend observed for the bulk structure. This implies that despite its different appearance and potential formation dynamics, the stabilisation of the surface layer is likely similar to that of the liquid domains in other regions of the bijel film. However, it also means that the wetting layer becomes increasingly difficult to distinguish for higher nanoparticle loadings, being practically invisible in the final two structures in Figure~\ref{NanoparticleLoading}. Consequently, although higher nanoparticle loadings yield better-defined bijel films, lower weight fractions of nanoparticles are more suitable for resolving the structure of wetting layers and the associated influence of substrate wettability. 

\subsection{Influence of Substrate Wettability}
\begin{figure*}
    \centering
    \includegraphics[width=\linewidth]{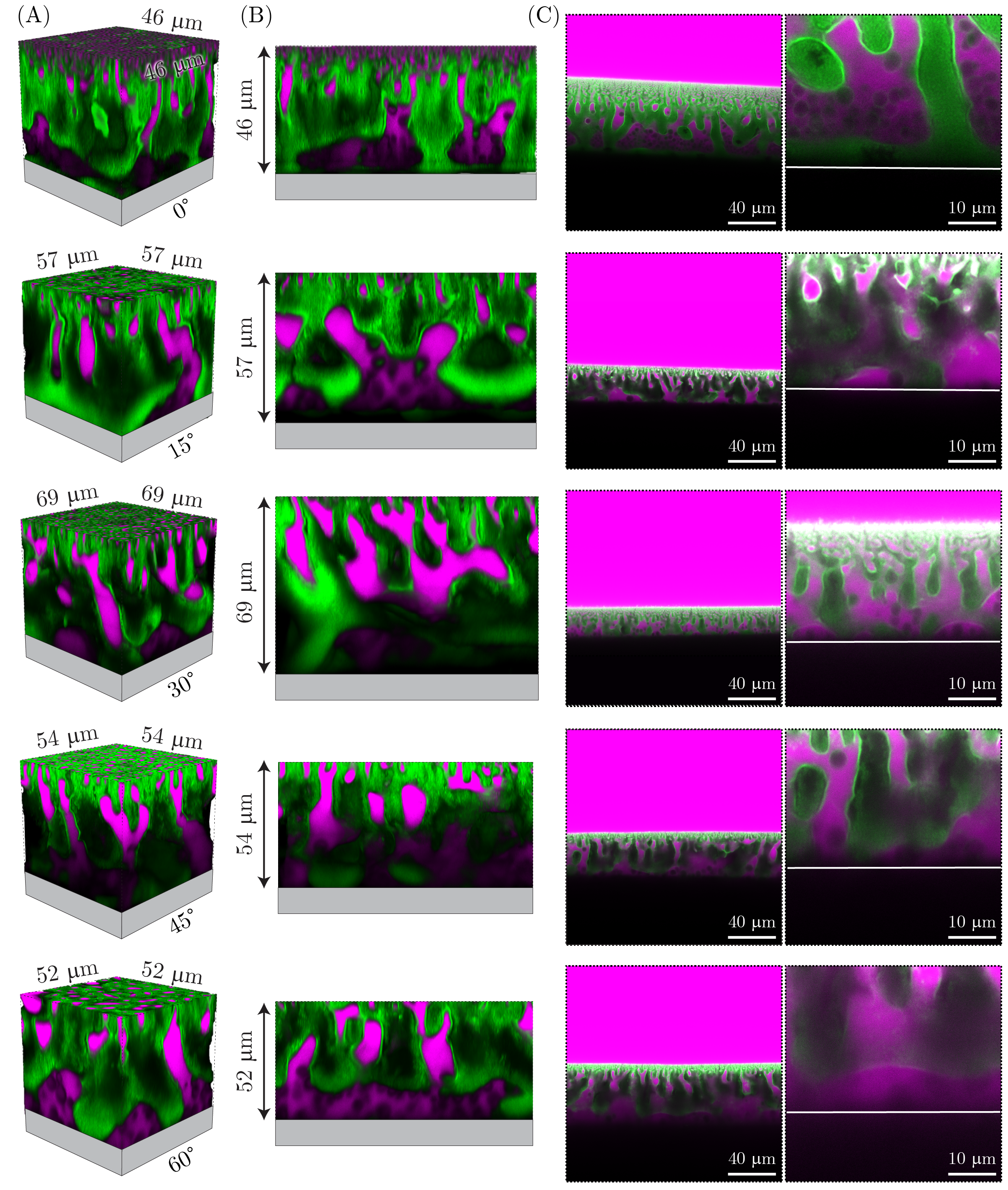}
    \caption{\textbf{(A)} 3D morphologies of supported bijel films, fabricated on glass substrates with varying wettability as indicated by the contact angle. All images in the same row share the same contact angle, ranging between \qtyrange{0}{60}{\degree}, with higher values corresponding to more hydrophobic surfaces. \textbf{(B)} Vertical cross-sections of the supported bijel films in (A), highlighting the shift in both composition and structure of the wetting layers at the bijel-substrate interface. \textbf{(C)} Complementary cross-sections of other supported bijel films, analogous to those in (A) and (B), acquired with the \textit{in situ} confocal imaging method. The right image zooms in on the bijel-substrate interface, which is indicated by the white line.}
    \label{SubstrateInfluence} 
\end{figure*}
With the insights from previous sections, supported bijel films were fabricated on glass substrates of varying wettability to investigate the nature of their surface layers. The weight fraction of the nanoparticles in these bijel films was kept at \qty{23}{w\percent}, sufficiently high to consistently produce a stable structure, but low enough for easy identification of the wetting layer. The contact angles of the glass slides varied between \qty{0}{\degree}, reflecting complete wetting, and \qty{60}{\degree}. Although bijel films were prepared on more hydrophobic surfaces with contact angles larger than \qty{60}{\degree}, these invariably exhibited complete detachment and could not be imaged together with the substrate. Consequently, the \qtyrange{0}{60}{\degree} range of contact angles effectively covered the full spectrum of surface hydrophilicity accessible under these experimental conditions.

The supported bijel films were subsequently imaged with confocal microscopy, the results of which are presented in Figure~\ref{SubstrateInfluence}. The content of Figure~\ref{SubstrateInfluence} is conceptually analogous to Figure~\ref{NanoparticleLoading}: Figure~\ref{SubstrateInfluence} (A) and Figure~\ref{SubstrateInfluence} (B) show 3D reconstructions of the bijel structure for different contact angles of the substrate, while Figure~\ref{SubstrateInfluence} (C) provides complementary vertical cross-sections acquired with the \textit{in situ} imaging method. 

At first glance, the 3D representations of the bijel films in Figure~\ref{SubstrateInfluence} (A) appear strongly similar between the differently modified substrates. In line with the phase-field simulations in Figure~\ref{PFSimulations}, this indicates that the bulk morphology of the bijel remains mostly unaffected by the properties of the surface. However, the cross-sectional images in Figure~\ref{SubstrateInfluence} (B) and (C) reveal considerable differences in the region near the bijel-substrate interface that can be attributed to the directing influence of the surface. In particular, there are noticeable trends in both the local structure and composition with increasing substrate hydrophobicity. 

As already shown in Figure~\ref{NanoparticleLoading}, the most hydrophilic substrate induces the formation of a distinct wetting layer consisting only of the water-rich phase. For more hydrophobic substrates, reflected in their higher contact angles, the region of the bijel adjacent to the surface becomes increasingly enriched with the oil-rich phase. The composition of the wetting layer eventually undergoes inversion, being predominantly oil-rich for the substrate with a contact angle of \qty{60}{\degree}. 

In addition to the composition, the structure of the surface layer also changes with the hydrophobicity of the substrate. Interestingly, the wetting layer only appears fully laminar for the most hydrophilic of the different substrates. Even a slight increase in surface hydrophobicity, such as going from a contact angle of \qty{0}{\degree} to \qty{15}{\degree}, seems sufficient to disrupt this laminar structure. Rather, the surface layer transitions into a more patch-like morphology, where both the oil-rich and water-rich phase are in contact with the substrate. The relative prevalence of these patches shifts towards the oil-rich phase as the substrates become increasingly hydrophobic, there only being a few water-rich patches for a contact angle of \qty{60}{\degree}. 

So far, the presented trends in both the structure and composition of the surface layers agree with the predictions from phase-field simulations. However, there is one notable exception: even the most hydrophobic substrates in Figure~\ref{SubstrateInfluence} do not result in a fully laminar wetting layer of the oil-rich phase. This observation makes sense when considering the adhesion behaviour of the bijel film. Given that the ambient liquid here is also an oil, the formation of a continuous layer of oil between the bijel and the substrate would simply cause the film to detach. With the confocal setup used in this work, the detachment of the bijel film effectively excludes these types of samples from being imaged together with the substrate. 

These results also help rationalise findings from a previous study on bijel formation using roll-to-roll STrIPS, where the attachment of the bijel films strongly depended on substrate wettability \cite{Siegel2025}. In that case, the conditions were effectively reversed: the bijel film was formed in an ambient phase consisting of water, instead of oil. Consequently, a more hydrophobic substrate would be required to induce the formation of oil-rich wetting layers and ensure the adhesion of the bijel film. This exactly matches the described results of the study, in which relatively hydrophilic substrates caused the bijel film to detach while it remained attached for more hydrophobic ones. 

\begin{figure*}
    \centering
    \includegraphics[width=\linewidth]{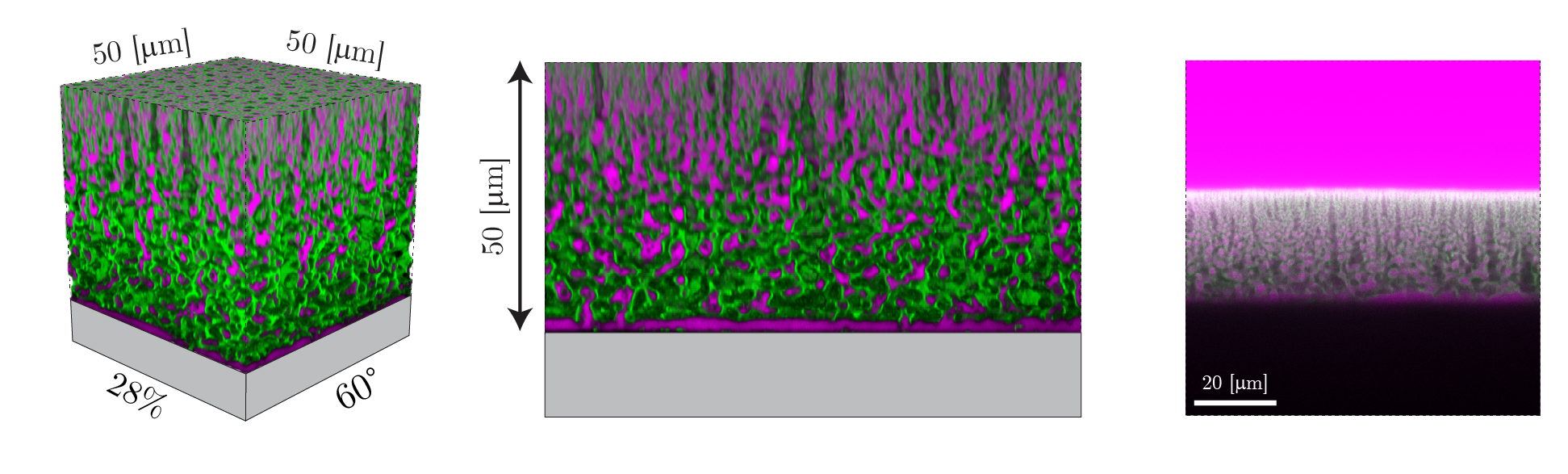}
    \caption{Confocal visualisation of a supported bijel film with a nanoparticle content of \qty{28}{w\percent}, coated on a comparatively hydrophobic substrate with a contact angle of \qty{60}{\degree}. Both the 3D structure on the left and its vertical cross-section in the middle reveal a planar, oil-rich wetting layer at the bijel-substrate interface. The complementary cross-section of a similar bijel film, shown on the right, highlights that such laminar, oil-rich wetting layers exist only locally and do not cover the full surface of the substrate.}
    \label{HydrophobicSubstrate} 
\end{figure*}

Although the oil-rich phase does not form a uniformly planar layer along the full extent of the substrate, some bijel films did exhibit a notably laminar surface structure on a local level. An example is shown in Figure~\ref{HydrophobicSubstrate}, which depicts a bijel film with a nanoparticle weight fraction of \qty{28}{w\percent} coated on a relatively hydrophobic glass slide with a contact angle of \qty{60}{\degree}. In both the 3D reconstruction on the left and its vertical cross-section in the centre, a laminar oil-rich layer is visible directly adjacent to the substrate. Consisting of two alternating layers of the oil- and water-rich phases, the general morphology of the surface region strongly resembles that of the phase-field simulation for a very hydrophobic surface in Figure~\ref{PFSimulations}. Moreover, the higher nanoparticle loading results in a considerably thinner wetting layer than for the bijel films in Figure~\ref{SubstrateInfluence}, which is fully in line with the findings in Figure~\ref{NanoparticleLoading}. Finally, the image on the right shows a complementary cross-section of an analogously produced bijel film with a similar structure. By imaging the water-rich patches around the more laminar oil layers, it is meant to emphasise the local nature of the latter. 

\section{Discussion}
The results of the previous sections univocally demonstrate the existence of wetting layers for STrIPS bijels, linking the composition and structure to the surface properties of the support material. Despite the consistency between simulation and experiment, the current methodology still leaves room for further refinement. 

Most importantly, the findings presented in this work remain largely qualitative. That is, the analysis of the supported bijel film is based mainly on the interpretation of confocal images. This type of analysis is well-suited for providing an initial indication of morphological trends, yet further substantiation of the observed behaviour requires a more quantitative approach. A particularly promising strategy would be the construction of composition-depth profiles, which show how an averaged measure of the local composition varies with depth in the bijel film. These profiles simultaneously characterise the nature of the wetting layer, through its composition and thickness, as well as the extent of the directing influence of the substrate into the bijel structure. The latter property could be quantitatively captured in terms of a correlation length, which is particularly relevant information for the application of thin-film bijels as templates for functional materials.

In principle, such profiles could already be extracted from the microscopy data presented here. Since both oil- and water-rich phases should be fully distinguishable in a confocal image, each horizontal cross-section in $z-$stack provides a measure of the local oil/water ratio at a certain depth in the bijel film. However, despite considerable optimisation efforts, the practical limitations of the current confocal dataset make it unsuitable for this purpose. The most prominent issue is the decay of fluorescent signal with depth in the bijel structure, which disproportionally affects the detection of the oil-rich phase compared to the water-rich phase. Consequently, the contributions of the water-rich phase become artificially inflated with sample depth, which is particularly problematic for thicker bijel films. In addition, the fabricated bijel films differ in thickness, and thus in the relative depth of the bijel-substrate interface. Together, these factors make it difficult to obtain an accurate quantitative characterisation of the wetting layer that is consistent between different samples. 

Fortunately, these problems also have relatively straightforward solutions for future work. The loss of fluorescent signal can be mitigated by further increasing the optical transparency of the sample, which for a bijel corresponds to better matching of the refractive indices between its liquid and solid phases. By using slot-die coating rather than dip-coating during the production of the bijel films, a uniform and constant thickness can be ensured. This investigation would also benefit from a complementary analysis method to independently verify the results found with confocal microscopy. Scanning electron microscopy, potentially combined with a focused ion beam as a confocal $z-$stack analogue, could hold particular promise.  

\section{Conclusions}
Combining numerical simulation with confocal microscopy, the study demonstrated the emergence of wetting layers in supported STrIPS bijels. The role of substrate wettability was investigated by fabricating supported bijel films on glass slides of varying hydrophobicity, achieved through controlled silanisation of their silica surfaces. The morphologies of these supported films were subsequently analysed with confocal microscopy, using a complementary dual-plane imaging approach to accurately resolve the full extent of the bijel structure. 

Confocal imaging revealed the presence of distinct wetting layers at the bijel-substrate interface.
As with the bulk domains, the thickness of these wetting layers could be tuned through the nanoparticle weight fraction in the bijel precursor. Moreover, both the composition and the general structure of the wetting layers strongly depended on the surface properties of the substrate. Increasing the substrate hydrophobicity induced a transition from a fully laminar, water-rich wetting layer to a more patch-like morphology of increasingly oil-rich character. 

Together, these findings provide vital insights into the directing influence of the supporting substrate on the bijel structure. As such, this understanding will be highly relevant for the successful production and implementation of supported bijel films as applied functional materials. 

\section*{Data Availability Statement}
The data that support the findings of this study are available from the corresponding author upon reasonable request.

\begin{acknowledgments}
This publication is part of the project ‘‘Bijel templated membranes for molecular separations’’ (with project number 18632 of the research programme Vidi 2019), which is financed by the Dutch Research Council (NWO).
\end{acknowledgments}

\section*{Author Contributions}
\textbf{Jesse M. Steenhoff:} Conceptualisation; Formal Analysis, Investigation; Methodology; Software; Validation; Visualisation; Writing - Original Draft; Writing- Review \& Editing. \textbf{Martin F. Haase:} Conceptualisation; Funding Acquisition; Supervision; Writing- Review \& Editing. 

\bibliography{Preprint}

@article{Banerjee2025,
   author = {Banerjee, Aihik and Khanal, Anjana and Okoro, Prince D. and Kharal, Shankar P. and Dalsania, Kevin and Kanjilal, Baishali and Iragavarapu, Shiril B. and Chen, Yiqing and Unagolla, Janitha M. and Liu, Huinan H. and Morgan, Joshua T. and Hesketh, Robert P. and Pezhouman, Arash and Ardehali, Reza and Anvari, Bahman and Haase, Martin F. and Noshadi, Iman},
   title = {Bicontinuous Interconnected Porous Biomaterials for Tissue Engineering and Regeneration},
   journal = {Small Science},
   volume = {5},
   number = {11},
   pages = {2500207},
   DOI = {https://doi.org/10.1002/smsc.202500207},
   url = {https://onlinelibrary.wiley.com/doi/abs/10.1002/smsc.202500207
https://onlinelibrary.wiley.com/doi/pdfdirect/10.1002/smsc.202500207?download=true},
   year = {2025},
   type = {Journal Article}
}

@article{Beunen2026,
   author = {Beunen, Johannes Martinus Peter and Harting, Jens},
   title = {Performance optimization of bijels as a novel type of catalyst support structure},
   journal = {Materials Horizons},
   ISSN = {2051-6347},
   DOI = {10.1039/D5MH01726B},
   url = {http://dx.doi.org/10.1039/D5MH01726B
https://pubs.rsc.org/en/Content/ArticleLanding/2026/MH/D5MH01726B
https://pubs.rsc.org/en/content/articlepdf/2026/mh/d5mh01726b},
   year = {2026},
   type = {Journal Article}
}

@article{Das2020,
   author = {Das, Prasenjit and Jaiswal, Prabhat K. and Puri, Sanjay},
   title = {Surface-directed spinodal decomposition on chemically patterned substrates},
   journal = {Physical Review E},
   volume = {102},
   number = {1},
   pages = {012803},
   note = {},
   DOI = {10.1103/PhysRevE.102.012803},
   url = {https://link.aps.org/doi/10.1103/PhysRevE.102.012803
https://journals.aps.org/pre/pdf/10.1103/PhysRevE.102.012803},
   year = {2020},
   type = {Journal Article}
}

@article{Flinn1994,
   author = {Flinn, D. H. and Guzonas, D. A. and Yoon, R. H.},
   title = {Characterization of silica surfaces hydrophobized by octadecyltrichlorosilane},
   journal = {Colloids and Surfaces A: Physicochemical and Engineering Aspects},
   volume = {87},
   number = {3},
   pages = {163–176},
   ISSN = {0927-7757},
   DOI = {https://doi.org/10.1016/0927-7757(94)80065-0},
   url = {https://www.sciencedirect.com/science/article/pii/0927775794800650},
   year = {1994},
   type = {Journal Article}
}

@article{Geoghegan2003,
   author = {Geoghegan, Mark and Krausch, Georg},
   title = {Wetting at polymer surfaces and interfaces},
   journal = {Progress in Polymer Science},
   volume = {28},
   number = {2},
   pages = {261–302},
   ISSN = {0079-6700},
   DOI = {https://doi.org/10.1016/S0079-6700(02)00080-1},
   url = {https://www.sciencedirect.com/science/article/pii/S0079670002000801},
   year = {2003},
   type = {Journal Article}
}

@article{Groisman2026,
   author = {Groisman, Luciano and Thorson, Todd J. and Botvinick, Elliot L. and Mohraz, Ali},
   title = {Hydraulic permeability of bijel-derived porous materials: Morphological origins of size-dominated flow},
   journal = {Journal of Membrane Science},
   volume = {742},
   pages = {125160},
   ISSN = {0376-7388},
   DOI = {https://doi.org/10.1016/j.memsci.2026.125160},
   url = {https://www.sciencedirect.com/science/article/pii/S0376738826000402
https://www.sciencedirect.com/science/article/pii/S0376738826000402?via%3Dihub},
   year = {2026},
   type = {Journal Article}
}

@article{Guo2016,
   author = {Guo, Xianwei and Han, Jiuhui and Liu, Pan and Chen, Luyang and Ito, Yoshikazu and Jian, Zelang and Jin, Tienan and Hirata, Akihiko and Li, Fujun and Fujita, Takeshi and Asao, Naoki and Zhou, Haoshen and Chen, Mingwei},
   title = {Hierarchical nanoporosity enhanced reversible capacity of bicontinuous nanoporous metal based Li-O2 battery},
   journal = {Scientific Reports},
   volume = {6},
   number = {1},
   pages = {33466},
   ISSN = {2045-2322},
   DOI = {10.1038/srep33466},
   url = {https://doi.org/10.1038/srep33466
https://www.nature.com/articles/srep33466.pdf},
   year = {2016},
   type = {Journal Article}
}

@article{Haase2015,
   author = {Haase, Martin F. and Stebe, Kathleen J. and Lee, Daeyeon},
   title = {Continuous Fabrication of Hierarchical and Asymmetric Bijel Microparticles, Fibers, and Membranes by Solvent Transfer-Induced Phase Separation (STRIPS)},
   journal = {Advanced Materials},
   volume = {27},
   number = {44},
   pages = {7065–7071},
   ISSN = {0935-9648},
   DOI = {https://doi.org/10.1002/adma.201503509},
   url = {https://advanced.onlinelibrary.wiley.com/doi/abs/10.1002/adma.201503509
https://advanced.onlinelibrary.wiley.com/doi/pdfdirect/10.1002/adma.201503509?download=true},
   year = {2015},
   type = {Journal Article}
}

@article{Tanaka2001,
   author = {Hajime, Tanaka},
   title = {Interplay between wetting and phase separation in binary fluid mixtures: roles of hydrodynamics},
   journal = {Journal of Physics: Condensed Matter},
   volume = {13},
   number = {21},
   pages = {4637},
   ISSN = {0953-8984},
   DOI = {10.1088/0953-8984/13/21/303},
   url = {https://doi.org/10.1088/0953-8984/13/21/303
https://iopscience.iop.org/article/10.1088/0953-8984/13/21/303},
   year = {2001},
   type = {Journal Article}
}

@article{Han2023,
   author = {Han, Junghun and Lee, Michael J. and Lee, Kyungbin and Lee, Young Jun and Kwon, Seung Ho and Min, Ju Hong and Lee, Eunji and Lee, Wonho and Lee, Seung Woo and Kim, Bumjoon J.},
   title = {Role of Bicontinuous Structure in Elastomeric Electrolytes for High-Energy Solid-State Lithium-Metal Batteries},
   journal = {Advanced Materials},
   volume = {35},
   number = {1},
   pages = {2205194},
   ISSN = {0935-9648},
   DOI = {https://doi.org/10.1002/adma.202205194},
   url = {https://advanced.onlinelibrary.wiley.com/doi/abs/10.1002/adma.202205194
https://advanced.onlinelibrary.wiley.com/doi/pdfdirect/10.1002/adma.202205194?download=true},
   year = {2023},
   type = {Journal Article}
}

@article{Hasan2016,
   author = {Hasan, Abshar and Pandey, Lalit M.},
   title = {Kinetic studies of attachment and re-orientation of octyltriethoxysilane for formation of self-assembled monolayer on a silica substrate},
   journal = {Materials Science and Engineering: C},
   volume = {68},
   pages = {423–429},
   ISSN = {0928-4931},
   DOI = {https://doi.org/10.1016/j.msec.2016.06.003},
   url = {https://www.sciencedirect.com/science/article/pii/S0928493116305707
https://www.sciencedirect.com/science/article/pii/S0928493116305707?via%3Dihub},
   year = {2016},
   type = {Journal Article}
}

@article{Jinnai2003,
   author = {Jinnai, Hiroshi and Kitagishi, Hitoshi and Hamano, Kazuki and Nishikawa, Yukihiro and Takahashi, Masaoki},
   title = {Effect of confinement on phase-separation processes in a polymer blend observed by laser scanning confocal microscopy},
   journal = {Physical Review E},
   volume = {67},
   number = {2},
   pages = {021801},
   note = {},
   DOI = {10.1103/PhysRevE.67.021801},
   url = {https://link.aps.org/doi/10.1103/PhysRevE.67.021801
https://journals.aps.org/pre/pdf/10.1103/PhysRevE.67.021801},
   year = {2003},
   type = {Journal Article}
}

@article{Jones1991,
   author = {Jones, Richard A. L. and Norton, Laura J. and Kramer, Edward J. and Bates, Frank S. and Wiltzius, Pierre},
   title = {Surface-directed spinodal decomposition},
   journal = {Physical Review Letters},
   volume = {66},
   number = {10},
   pages = {1326–1329},
   note = {},
   DOI = {10.1103/PhysRevLett.66.1326},
   url = {https://link.aps.org/doi/10.1103/PhysRevLett.66.1326
https://journals.aps.org/prl/pdf/10.1103/PhysRevLett.66.1326},
   year = {1991},
   type = {Journal Article}
}

@article{Khan2022,
   author = {Khan, Mohd A. and Sprockel, Alessio J. and Macmillan, Katherine A. and Alting, Meyer T. and Kharal, Shankar P. and Boakye-Ansah, Stephen and Haase, Martin F.},
   title = {Nanostructured, Fluid-Bicontinuous Gels for Continuous-Flow Liquid–Liquid Extraction},
   journal = {Advanced Materials},
   volume = {34},
   number = {18},
   pages = {2109547},
   ISSN = {0935-9648},
   DOI = {https://doi.org/10.1002/adma.202109547},
   url = {https://advanced.onlinelibrary.wiley.com/doi/abs/10.1002/adma.202109547
https://advanced.onlinelibrary.wiley.com/doi/pdfdirect/10.1002/adma.202109547?download=true},
   year = {2022},
   type = {Journal Article}
}

@article{Krausch1995,
   author = {Krausch, Georg},
   title = {Surface induced self assembly in thin polymer films},
   journal = {Materials Science and Engineering: R: Reports},
   volume = {14},
   number = {1},
   pages = {v–94},
   ISSN = {0927-796X},
   DOI = {https://doi.org/10.1016/0927-796X(94)00173-1},
   url = {https://www.sciencedirect.com/science/article/pii/0927796X94001731},
   year = {1995},
   type = {Journal Article}
}

@article{Li2020,
   author = {Li, Qian and Chen, Chuanshuang and Li, Chen and Liu, Ruiyi and Bi, Shuai and Zhang, Pengfei and Zhou, Yongfeng and Mai, Yiyong},
   title = {Ordered Bicontinuous Mesoporous Polymeric Semiconductor Photocatalyst},
   journal = {ACS Nano},
   volume = {14},
   number = {10},
   pages = {13652–13662},
   note = {},
   ISSN = {1936-0851},
   DOI = {10.1021/acsnano.0c05797},
   url = {https://doi.org/10.1021/acsnano.0c05797
https://pubs.acs.org/doi/pdf/10.1021/acsnano.0c05797?ref=article_openPDF},
   year = {2020},
   type = {Journal Article}
}

@article{Lin1994,
   author = {Lin, M. Y. and Sinha, S. K. and Drake, J. M. and Wu, X. l and Thiyagarajan, P. and Stanley, H. B.},
   title = {Study of phase separation of a binary fluid mixture in confined geometry},
   journal = {Physical Review Letters},
   volume = {72},
   number = {14},
   pages = {2207–2210},
   note = {},
   DOI = {10.1103/PhysRevLett.72.2207},
   url = {https://link.aps.org/doi/10.1103/PhysRevLett.72.2207
https://journals.aps.org/prl/pdf/10.1103/PhysRevLett.72.2207},
   year = {1994},
   type = {Journal Article}
}

@article{Lowe2011,
   author = {Lowe, Randall D. and Pellow, Matthew A. and Stack, T. Daniel P. and Chidsey, Christopher E. D.},
   title = {Deposition of Dense Siloxane Monolayers from Water and Trimethoxyorganosilane Vapor},
   journal = {Langmuir},
   volume = {27},
   number = {16},
   pages = {9928–9935},
   note = {},
   ISSN = {0743-7463},
   DOI = {10.1021/la201333y},
   url = {https://doi.org/10.1021/la201333y
https://pubs.acs.org/doi/pdf/10.1021/la201333y?ref=article_openPDF},
   year = {2011},
   type = {Journal Article}
}

@article{Moffitt2002,
   author = {Moffitt, Matthew and Rharbi, Yahya and Li, Huxi and Winnik, Mitchell A.},
   title = {Novel Morphology Evolution in Thick Films of a Polymer Blend},
   journal = {Macromolecules},
   volume = {35},
   number = {9},
   pages = {3321–3324},
   note = {},
   ISSN = {0024-9297},
   DOI = {10.1021/ma011679+},
   url = {https://doi.org/10.1021/ma011679+
https://pubs.acs.org/doi/pdf/10.1021/ma011679%2B?ref=article_openPDF},
   year = {2002},
   type = {Journal Article}
}

@article{Okoro2026,
   author = {Okoro, Prince D. and Dalsania, Kevin and Iragavarapu, Shiril B. and Dela Cruz, Benjamin and Banerjee, Aihik and Basaranbilek, Merve and Haase, Martin F. and Anvari, Bahman and Noshadi, Iman},
   title = {Bicontinuous Microarchitected Scaffolds Provide Topographic Cues That Govern Neuronal Behavior and Maturation},
   journal = {Advanced Functional Materials},
   volume = {36},
   number = {5},
   pages = {e09452},
   ISSN = {1616-301X},
   DOI = {https://doi.org/10.1002/adfm.202509452},
   url = {https://advanced.onlinelibrary.wiley.com/doi/abs/10.1002/adfm.202509452
https://advanced.onlinelibrary.wiley.com/doi/pdfdirect/10.1002/adfm.202509452?download=true},
   year = {2026},
   type = {Journal Article}
}

@article{Siegel2024,
   author = {Siegel, Henrik and de Ruiter, Mariska and Athanasiou, Georgios and Hesseling, Cos M. and Haase, Martin F.},
   title = {Roll-to-Roll Fabrication of Bijels via Solvent Transfer Induced Phase Separation (R2R-STrIPS)},
   journal = {Advanced Materials Technologies},
   volume = {9},
   number = {3},
   pages = {2301525},
   ISSN = {2365-709X},
   DOI = {https://doi.org/10.1002/admt.202301525},
   url = {https://doi.org/10.1002/admt.202301525
https://advanced.onlinelibrary.wiley.com/doi/pdfdirect/10.1002/admt.202301525?download=true},
   year = {2024},
   type = {Journal Article}
}

@article{Siegel2025,
   author = {Siegel, Henrik and Haase, Martin F.},
   title = {Bijel Membranes with Tunable Porosity for pH-responsive Microfiltration},
   journal = {Small},
   volume = {21},
   number = {36},
   pages = {2504768},
   ISSN = {1613-6810},
   DOI = {https://doi.org/10.1002/smll.202504768},
   url = {https://onlinelibrary.wiley.com/doi/abs/10.1002/smll.202504768
https://onlinelibrary.wiley.com/doi/pdfdirect/10.1002/smll.202504768?download=true},
   year = {2025},
   type = {Journal Article}
}

@article{Siegel2022,
   author = {Siegel, Henrik and Sprockel, Alessio J. and Schwenger, Matthew S. and Steenhoff, Jesse M. and Achterhuis, Iske and de Vos, Wiebe M. and Haase, Martin F.},
   title = {Synthesis and Polyelectrolyte Functionalization of Hollow Fiber Membranes Formed by Solvent Transfer Induced Phase Separation},
   journal = {ACS Applied Materials \& Interfaces},
   volume = {14},
   number = {38},
   pages = {43195–43206},
   note = {},
   ISSN = {1944-8244},
   DOI = {10.1021/acsami.2c10343},
   url = {https://doi.org/10.1021/acsami.2c10343
https://pubs.acs.org/doi/pdf/10.1021/acsami.2c10343?ref=article_openPDF},
   year = {2022},
   type = {Journal Article}
}

@article{Tanaka1993,
   author = {Tanaka, H.},
   title = {Interplay between Phase Separation and Wetting for a Polymer Mixture Confined in a Two-Dimensional Capillary: Wetting-Induced Domain Ordering and Coarsening},
   journal = {Europhysics Letters},
   volume = {24},
   number = {8},
   pages = {665},
   ISSN = {0295-5075},
   DOI = {10.1209/0295-5075/24/8/008},
   url = {https://doi.org/10.1209/0295-5075/24/8/008},
   year = {1993},
   type = {Journal Article}
}

@article{Tanaka1993_2,
   author = {Tanaka, Hajime},
   title = {Wetting dynamics in a confined symmetric binary mixture undergoing phase separation},
   journal = {Physical Review Letters},
   volume = {70},
   number = {18},
   pages = {2770–2773},
   note = {},
   DOI = {10.1103/PhysRevLett.70.2770},
   url = {https://link.aps.org/doi/10.1103/PhysRevLett.70.2770
https://journals.aps.org/prl/pdf/10.1103/PhysRevLett.70.2770},
   year = {1993},
   type = {Journal Article}
}

@article{Wang2000,
   author = {Wang, Howard and Composto, Russell J.},
   title = {Thin film polymer blends undergoing phase separation and wetting: Identification of early, intermediate, and late stages},
   journal = {The Journal of Chemical Physics},
   volume = {113},
   number = {22},
   pages = {10386–10397},
   ISSN = {0021-9606},
   DOI = {10.1063/1.1322638},
   url = {https://doi.org/10.1063/1.1322638
https://watermark02.silverchair.com/10386_1_online.pdf?token=AQECAHi208BE49Ooan9kkhW_Ercy7Dm3ZL_9Cf3qfKAc485ysgAABY4wggWKBgkqhkiG9w0BBwagggV7MIIFdwIBADCCBXAGCSqGSIb3DQEHATAeBglghkgBZQMEAS4wEQQMrz0iVcMSiE9OlGTZAgEQgIIFQcNkIgnoohF6BUITqR5JYco85MbKEVP7zlNZViKnEVzxZtsiJy0Q7AyqPwIhiS6uwrixUOdQze-dt3Ru4ufL-YhKD-vzGqEu0UveKjVXXXod8dIIn_G-OX479D57fMd-WsqcPmKRT7nYu1oob3OaxD_ghdbTX1KM5gi1hjVqkGXqYboYyon_l6ci1aZ0_9nlU8ta07PCnZtHfUIpLi1hDFmztZV1leLf5JwSuobrhoqfNSENcII7gIXX640ujmWtZD1S40OirrcPlioyFUhNEPgtK2adXxonlXGCP3d1Rkznc9nVl_0ZSH5HPIJRj2lV0fxwwhvSAsH_C-CiZqEayaWQQDYeydnVy1EW51_rllYMq7HuiZcbbs6jZD1ZK9EVakHDiMTKMweeR_HJDfd1zwsUEEbWIs38qKg0-rkkANuVu_qpo9Cte1LYVYlaaZ3gLronU8vMbFKMaze3gjqlINpYhyjg4_BNr9BIGDyxHFFyRtA0pQjBddMOTuYZ5tcGoLc50kMXtd4EC6tsHaRb2xz1_TAirPqm5uMv_6nzjp_82MGeywSK4zvIVvPNcV9-pmAF6hCh7-jln6kbsQmnZ_qEyOWWbWtsyPllZKPnvFuGJKNP4cYqMUJ9z1MEnjYOp9_cVfRMGbp1c_KudRfrXtu0Db3d7Ffg8fGxnzYpEdDi2w23VcOYs1Y8NGpyBQpwXwVkOm3zTK2z3ETqJufdKkhsooxcFrX2JBoC6dR_tFWcUVwazAoLjxxjdkrmuYucYd8FTlzivXfrBy-z2TJ-cq5UDrTaN4UWdlwQSGCDJT3oQ_7T0i6CcTIzFRxueqaavwNmoM-0bvzGpJ3m_bVpZlnQ1cdNHCvD_VloFTxc2lPb6XEtKlixK8Loek6cXaQk_UvoI1-sL53uW3K7hKn9CNFQqbhDkYxVAwstzpY1aw_6v-DNATKNz-tkPKiBnV_jH9U96HZkmkKLilMsUTXaA4BQlHepJAZHVOutUzabFnL6bZDi9OpcdbXrh1lsRl9r2hlmtoZVfFhROR6_ZG4JpuoMMIQDKnQ0oU69pHxcw3NfyNJbJ9U2rj2LLiq5ljLbYNFZX7Cv8WZXasdkC-zkUzVHC4z544odjE0tbJEE05WAugHvwf9LlWhHx9jtHnKPslZS1bxqG3k-Rv7LjokB-DRddccIK5JkUr9VFfXc4fFgJm8-ngTR36sGuFeOOrMUP-Lgba98oLCioV4Y8J_5YAquT1y6fs0t7Y1THJaLL6OPsuuI1NtoFsU6iEve3Dh6o9ltFIGvi3JFAkxthNxAr71PyONDw9WNDVUQBzoYQBxB950H2EyPSRd_AKcjkfoo3M9a2EeIEsbAx2mgensO1LBX7cbg7lSr4Ki0FNBlQ_AN_oFGTORMmfsyqzctFziWyPNMods8cVtb9pJFkkZ-7eKCEOcszogLqxFdrhcF1mdoFX7apKjwvt14g8gosTOFS3qxH58VzSQZIGZe2Iwrwx4FnO-O4yxWWBoB3eEqII1eGucjJLO9CnMPlHjVduLYAJf5s2pWKjDreiHe7R57GZpf4PyPDJpUp5AOfvCLS0T0VDjUzm8qoMF3gNackZGt6bGCBQOIR0-QwdvZh9gIvDiiX285dRp6sfYZcEBsGOgxWWxj_WEh_k5f_weqS-mpL9OnG-0SyECB2B-Bmbr49TQKlHh4RqeqFDsTp3XUpzOkWVU01SsWwieAxI_B58_Pctsq0SOfrr2Lh53awEPOi9jIrwYvbWjbPq1vsfJKiJj8ijYwokjcRboFXTn2JxAM3pY},
   year = {2000},
   type = {Journal Article}
}

@article{Xiang2023,
   author = {Xiang, Luoxing and Li, Qian and Li, Chen and Yang, Qiqi and Xu, Fugui and Mai, Yiyong},
   title = {Block Copolymer Self-Assembly Directed Synthesis of Porous Materials with Ordered Bicontinuous Structures and Their Potential Applications},
   journal = {Advanced Materials},
   volume = {35},
   number = {5},
   pages = {2207684},
   ISSN = {0935-9648},
   DOI = {https://doi.org/10.1002/adma.202207684},
   url = {https://advanced.onlinelibrary.wiley.com/doi/abs/10.1002/adma.202207684
https://advanced.onlinelibrary.wiley.com/doi/pdfdirect/10.1002/adma.202207684?download=true},
   year = {2023},
   type = {Journal Article}
}

@article{Zielasek2006,
   author = {Zielasek, Volkmar and Jürgens, Birte and Schulz, Christian and Biener, Jürgen and Biener, Monika M. and Hamza, Alex V. and Bäumer, Marcus},
   title = {Gold Catalysts: Nanoporous Gold Foams},
   journal = {Angewandte Chemie International Edition},
   volume = {45},
   number = {48},
   pages = {8241–8244},
   ISSN = {1433-7851},
   DOI = {https://doi.org/10.1002/anie.200602484},
   url = {https://onlinelibrary.wiley.com/doi/abs/10.1002/anie.200602484
https://onlinelibrary.wiley.com/doi/pdfdirect/10.1002/anie.200602484?download=true},
   year = {2006},
   type = {Journal Article}
}

@article{Scriven1976,
   author = {Scriven, L. E.},
   title = {Equilibrium bicontinuous structure},
   journal = {Nature},
   volume = {263},
   number = {5573},
   pages = {123–125},
   ISSN = {1476-4687},
   DOI = {10.1038/263123a0},
   url = {https://doi.org/10.1038/263123a0},
   year = {1976},
   type = {Journal Article}
}

@article{Ruiter2024,
   author = {de Ruiter, Mariska and Alting, Meyer T. and Siegel, Henrik and Haase, Martin F.},
   title = {Dual access to the fluid networks of colloid-stabilized bicontinuous emulsions through uninterrupted connections},
   journal = {Materials Horizons},
   volume = {11},
   number = {20},
   pages = {4987–4997},
   ISSN = {2051-6347},
   DOI = {10.1039/D4MH00495G},
   url = {http://dx.doi.org/10.1039/D4MH00495G
https://pubs.rsc.org/en/content/articlepdf/2024/mh/d4mh00495g},
   year = {2024},
   type = {Journal Article}
}

@article{Haase2017,
   author = {Haase, Martin F. and Jeon, Harim and Hough, Noah and Kim, Jong Hak and Stebe, Kathleen J. and Lee, Daeyeon},
   title = {Multifunctional nanocomposite hollow fiber membranes by solvent transfer induced phase separation},
   journal = {Nature Communications},
   volume = {8},
   number = {1},
   pages = {1234},
   ISSN = {2041-1723},
   DOI = {10.1038/s41467-017-01409-3},
   url = {https://doi.org/10.1038/s41467-017-01409-3
https://www.nature.com/articles/s41467-017-01409-3.pdf},
   year = {2017},
   type = {Journal Article}
}

@article{Cha2019,
   author = {Cha, Sanghak and Lim, Hyun Gyu and Haase, Martin F. and Stebe, Kathleen J. and Jung, Gyoo Yeol and Lee, Daeyeon},
   title = {Bicontinuous Interfacially Jammed Emulsion Gels (bijels) as Media for Enabling Enzymatic Reactive Separation of a Highly Water Insoluble Substrate},
   journal = {Scientific Reports},
   volume = {9},
   number = {1},
   pages = {6363},
   ISSN = {2045-2322},
   DOI = {10.1038/s41598-019-42769-8},
   url = {https://doi.org/10.1038/s41598-019-42769-8
https://www.nature.com/articles/s41598-019-42769-8.pdf},
   year = {2019},
   type = {Journal Article}
}

@article{Vitantonio2019,
   author = {Di Vitantonio, Giuseppe and Wang, Tiancheng and Haase, Martin F. and Stebe, Kathleen J. and Lee, Daeyeon},
   title = {Robust Bijels for Reactive Separation via Silica-Reinforced Nanoparticle Layers},
   journal = {ACS Nano},
   volume = {13},
   number = {1},
   pages = {26–31},
   note = {},
   ISSN = {1936-0851},
   DOI = {10.1021/acsnano.8b05718},
   url = {https://doi.org/10.1021/acsnano.8b05718
https://pubs.acs.org/doi/pdf/10.1021/acsnano.8b05718?ref=article_openPDF},
   year = {2019},
   type = {Journal Article}
}

@article{Herzig2007,
   author = {Herzig, E. M. and White, K. A. and Schofield, A. B. and Poon, W. C. K. and Clegg, P. S.},
   title = {Bicontinuous emulsions stabilized solely by colloidal particles},
   journal = {Nature Materials},
   volume = {6},
   number = {12},
   pages = {966–971},
   ISSN = {1476-4660},
   DOI = {10.1038/nmat2055},
   url = {https://doi.org/10.1038/nmat2055
https://www.nature.com/articles/nmat2055.pdf},
   year = {2007},
   type = {Journal Article}
}

@article{Stratford2005,
   author = {Stratford, K. and Adhikari, R. and Pagonabarraga, I. and Desplat, J.-C. and Cates, M. E.},
   title = {Colloidal Jamming at Interfaces: A Route to Fluid-Bicontinuous Gels},
   journal = {Science},
   volume = {309},
   number = {5744},
   pages = {2198–2201},
   DOI = {doi:10.1126/science.1116589},
   url = {https://www.science.org/doi/abs/10.1126/science.1116589
https://www.science.org/doi/pdf/10.1126/science.1116589?download=true},
   year = {2005},
   type = {Journal Article}
}

@article{Tavacoli2011,
   author = {Tavacoli, Joe W. and Thijssen, Job H. J. and Schofield, Andrew B. and Clegg, Paul S.},
   title = {Novel, Robust, and Versatile Bijels of Nitromethane, Ethanediol, and Colloidal Silica: Capsules, Sub-Ten-Micrometer Domains, and Mechanical Properties},
   journal = {Advanced Functional Materials},
   volume = {21},
   number = {11},
   pages = {2020–2027},
   ISSN = {1616-301X},
   DOI = {https://doi.org/10.1002/adfm.201002562},
   url = {https://advanced.onlinelibrary.wiley.com/doi/abs/10.1002/adfm.201002562},
   year = {2011},
   type = {Journal Article}
}

@article{Witt2013,
   author = {Witt, Jessica A. and Mumm, Daniel R. and Mohraz, Ali},
   title = {Bijel reinforcement by droplet bridging: a route to bicontinuous materials with large domains},
   journal = {Soft Matter},
   volume = {9},
   number = {29},
   pages = {6773–6780},
   ISSN = {1744-683X},
   DOI = {10.1039/C3SM00130J},
   url = {http://dx.doi.org/10.1039/C3SM00130J},
   year = {2013},
   type = {Journal Article}
}

@article{Steenhoff2026,
   author = {Eij, Elisabeth C. and de Graaf, Joost and Haase, Martin F. and Steenhoff, Jesse M.},
   title = {Phase-field models for particle-stabilized emulsions},
   journal = {The Journal of Chemical Physics},
   volume = {164},
   number = {12},
   ISSN = {0021-9606},
   DOI = {10.1063/5.0329556},
   url = {https://doi.org/10.1063/5.0329556},
   year = {2026},
   type = {Journal Article}
}

@phdthesis{Steenhoff2026_2,
    author = {Steenhoff, Jesse M.} ,
    title = {Solvent-Mediated Phase Separation for Particle-Stabilised Bicontinuous Emulsions} ,
    school = {Utrecht University} ,
    year = {2026}
}

@article{Werner2018,
   author = {Werner, J. G. and Rodríguez-Calero, G. G. and Abruña, H. D. and Wiesner, U.},
   title = {Block copolymer derived 3-D interpenetrating multifunctional gyroidal nanohybrids for electrical energy storage},
   journal = {Energy \& Environmental Science},
   volume = {11},
   number = {5},
   pages = {1261–1270},
   ISSN = {1754-5692},
   DOI = {10.1039/C7EE03571C},
   url = {http://dx.doi.org/10.1039/C7EE03571C},
   year = {2018},
   type = {Journal Article}
}

@article{Bai2026,
   author = {Bai, Shen-wei and Hao, Yong-jie and Yan, Yue-kai and Li, Lian-bi and Cheng, Peng-fei and Tu, Zhe-yan and Zhou, Hu and Cheng, Lin and Mei, Hui and Ding, Wei-guo},
   title = {Triply Periodic Minimal Surface (TPMS) Structures in 3D-Printed Catalyst Supports: Advanced Optimization of Heat Transfer, Mass Transport, and Mechanical Stability for Catalysis},
   journal = {Advanced Materials Technologies},
   volume = {n/a},
   number = {n/a},
   year = {2026},
   pages = {e02352},
   ISSN = {2365-709X},
   DOI = {https://doi.org/10.1002/admt.202502352},
   url = {https://advanced.onlinelibrary.wiley.com/doi/abs/10.1002/admt.202502352},
   type = {Journal Article}
}

@article{Dudaryeva2025,
   author = {Dudaryeva, Oksana Y. and Cousin, Lucien and Krajnovic, Leila and Gröbli, Gian and Sapkota, Virbin and Ritter, Lauritz and Deshmukh, Dhananjay and Cui, Yifan and Style, Robert W. and Levato, Riccardo and Labouesse, Céline and Tibbitt, Mark W.},
   title = {Tunable Bicontinuous Macroporous Cell Culture Scaffolds via Kinetically Controlled Phase Separation},
   journal = {Advanced Materials},
   volume = {37},
   number = {7},
   pages = {2410452},
   ISSN = {0935-9648},
   DOI = {https://doi.org/10.1002/adma.202410452},
   url = {https://advanced.onlinelibrary.wiley.com/doi/abs/10.1002/adma.202410452},
   year = {2025},
   type = {Journal Article}
}

@article{Hori2019,
   author = {Hori, Aruto and Watabe, Yuki and Yamada, Masumi and Yajima, Yuya and Utoh, Rie and Seki, Minoru},
   title = {One-Step Formation of Microporous Hydrogel Sponges Encapsulating Living Cells by Utilizing Bicontinuous Dispersion of Aqueous Polymer Solutions},
   journal = {ACS Applied Bio Materials},
   volume = {2},
   number = {5},
   pages = {2237–2245},
   note = {},
   DOI = {10.1021/acsabm.9b00194},
   url = {https://doi.org/10.1021/acsabm.9b00194},
   year = {2019},
   type = {Journal Article}
}

@article{Wiesner2023,
   author = {Wiesner, Florian and Limper, Alexander and Marth, Cedric and Brodersen, Anselm and Wessling, Matthias and Linkhorst, John},
   title = {Additive Manufacturing of Intertwined Electrode Pairs - Guided Mass Transport with Gyroids},
   journal = {Advanced Engineering Materials},
   volume = {25},
   number = {1},
   pages = {2200986},
   ISSN = {1438-1656},
   DOI = {https://doi.org/10.1002/adem.202200986},
   url = {https://advanced.onlinelibrary.wiley.com/doi/abs/10.1002/adem.202200986},
   year = {2023},
   type = {Journal Article}
}

@article{Hsueh2012,
   author = {Hsueh, Han-Yu and Ho, Rong-Ming},
   title = {Bicontinuous Ceramics with High Surface Area from Block Copolymer Templates},
   journal = {Langmuir},
   volume = {28},
   number = {22},
   pages = {8518–8529},
   note = {},
   ISSN = {0743-7463},
   DOI = {10.1021/la3009706},
   url = {https://doi.org/10.1021/la3009706},
   year = {2012},
   type = {Journal Article}
}

@article{Kwon2025,
   author = {Kwon, Yongmin and Choi, Su Bin and Kim, Minjung and Goo, Bon Seung and Lee, Young Wook and Hong, Jong Wook and Han, Sang Woo},
   title = {One-Pot Synthesis of Bicontinuous Palladium Nanocubes with Distinct Catalytic Properties for Various Electrocatalysis and Heterogeneous Catalysis},
   journal = {Nano Letters},
   volume = {25},
   number = {3},
   pages = {1226–1232},
   note = {},
   ISSN = {1530-6984},
   DOI = {10.1021/acs.nanolett.4c06396},
   url = {https://doi.org/10.1021/acs.nanolett.4c06396},
   year = {2025},
   type = {Journal Article}
}

@article{Huang2015,
   author = {Huang, Shao-Zhuan and Jin, Jun and Cai, Yi and Li, Yu and Deng, Zhao and Zeng, Jun-Yang and Liu, Jing and Wang, Chao and Hasan, Tawfique and Su, Bao-Lian},
   title = {Three-Dimensional (3D) Bicontinuous Hierarchically Porous Mn2O3 Single Crystals for High Performance Lithium-Ion Batteries},
   journal = {Scientific Reports},
   volume = {5},
   number = {1},
   pages = {14686},
   ISSN = {2045-2322},
   DOI = {10.1038/srep14686},
   url = {https://doi.org/10.1038/srep14686},
   year = {2015},
   type = {Journal Article}
}

@article{Sheng2021,
   author = {Sheng, Qingqing and Li, Qian and Xiang, Luoxing and Huang, Tao and Mai, Yiyong and Han, Lu},
   title = {Double diamond structured bicontinuous mesoporous titania templated by a block copolymer for anode material of lithium-ion battery},
   journal = {Nano Research},
   volume = {14},
   number = {4},
   pages = {992–997},
   ISSN = {1998-0000},
   DOI = {10.1007/s12274-020-3139-4},
   url = {https://doi.org/10.1007/s12274-020-3139-4},
   year = {2021},
   type = {Journal Article}
}

@article{Tang2024,
   author = {Tang, Chen and Lu, Wei and Zhang, Yixiao and Zhang, Wenwei and Cui, Congcong and Liu, Pan and Han, Lu and Qian, Xiaoshi and Chen, Liwei and Xu, Fugui and Mai, Yiyong},
   title = {Toward Ultrahigh Rate and Cycling Performance of Cathode Materials of Sodium Ion Battery by Introducing a Bicontinuous Porous Structure},
   journal = {Advanced Materials},
   volume = {36},
   number = {26},
   pages = {2402005},
   ISSN = {0935-9648},
   DOI = {https://doi.org/10.1002/adma.202402005},
   url = {https://advanced.onlinelibrary.wiley.com/doi/abs/10.1002/adma.202402005},
   year = {2024},
   type = {Journal Article}
}

@article{Pang2020,
   author = {Pang, Ruizhi and Chen, Kai K. and Han, Yang and Ho, W. S. Winston},
   title = {Highly permeable polyethersulfone substrates with bicontinuous structure for composite membranes in CO2/N2 separation},
   journal = {Journal of Membrane Science},
   volume = {612},
   pages = {118443},
   ISSN = {0376-7388},
   DOI = {https://doi.org/10.1016/j.memsci.2020.118443},
   url = {https://www.sciencedirect.com/science/article/pii/S037673882031019X},
   year = {2020},
   type = {Journal Article}
}

@article{Zhou2007,
   author = {Zhou, Meijuan and Nemade, Parag R. and Lu, Xiaoyun and Zeng, Xiaohui and Hatakeyama, Evan S. and Noble, Richard D. and Gin, Douglas L.},
   title = {New Type of Membrane Material for Water Desalination Based on a Cross-Linked Bicontinuous Cubic Lyotropic Liquid Crystal Assembly},
   journal = {Journal of the American Chemical Society},
   volume = {129},
   number = {31},
   pages = {9574–9575},
   note = {},
   ISSN = {0002-7863},
   DOI = {10.1021/ja073067w},
   url = {https://doi.org/10.1021/ja073067w},
   year = {2007},
   type = {Journal Article}
}

@article{Cates2008,
   author = {Cates, Michael E. and Clegg, Paul S.},
   title = {Bijels: a new class of soft materials},
   journal = {Soft Matter},
   volume = {4},
   number = {11},
   pages = {2132–2138},
   ISSN = {1744-683X},
   DOI = {10.1039/B807312K},
   url = {http://dx.doi.org/10.1039/B807312K
https://pubs.rsc.org/en/content/articlepdf/2008/sm/b807312k},
   year = {2008},
   type = {Journal Article}
}

@article{Bai2015,
   author = {Bai, Lian and Fruehwirth, John W. and Cheng, Xiang and Macosko, Christopher W.},
   title = {Dynamics and rheology of nonpolar bijels},
   journal = {Soft Matter},
   volume = {11},
   number = {26},
   pages = {5282-5293},
   ISSN = {1744-683X},
   DOI = {10.1039/C5SM00994D},
   url = {http://dx.doi.org/10.1039/C5SM00994D},
   year = {2015},
   type = {Journal Article}
}

@article{Cai2015,
   author = {Cai, Dongyu and Clegg, Paul S.},
   title = {Stabilizing bijels using a mixture of fumed silica nanoparticles},
   journal = {Chemical Communications},
   volume = {51},
   number = {95},
   pages = {16984-16987},
   ISSN = {1359-7345},
   DOI = {10.1039/C5CC07346D},
   url = {http://dx.doi.org/10.1039/C5CC07346D},
   year = {2015},
   type = {Journal Article}
}

@article{Sprockel2023,
   author = {Sprockel, Alessio J. and Khan, Mohd A. and de Ruiter, Mariska and Alting, Meyer T. and Macmillan, Katherine A. and Haase, Martin F.},
   title = {Fabrication of bijels with sub-micron domains via a single-channel flow device},
   journal = {Colloids and Surfaces A: Physicochemical and Engineering Aspects},
   volume = {666},
   pages = {131306},
   ISSN = {0927-7757},
   DOI = {https://doi.org/10.1016/j.colsurfa.2023.131306},
   url = {https://www.sciencedirect.com/science/article/pii/S0927775723003904
https://www.sciencedirect.com/science/article/pii/S0927775723003904?via%3Dihub},
   year = {2023},
   type = {Journal Article}
}

@article{Bruder1992,
   author = {Bruder, F. and Brenn, R.},
   title = {Spinodal decomposition in thin films of a polymer blend},
   journal = {Physical Review Letters},
   volume = {69},
   number = {4},
   pages = {624–627},
   note = {},
   DOI = {10.1103/PhysRevLett.69.624},
   url = {https://link.aps.org/doi/10.1103/PhysRevLett.69.624},
   year = {1992},
   type = {Journal Article}
}

@article{Geoghegan2000,
   author = {Geoghegan, Mark and Ermer, Hubert and Jüngst, Gerald and Krausch, Georg and Brenn, Rüdiger},
   title = {Wetting in a phase separating polymer blend film: Quench depth dependence},
   journal = {Physical Review E},
   volume = {62},
   number = {1},
   pages = {940–950},
   note = {},
   DOI = {10.1103/PhysRevE.62.940},
   url = {https://link.aps.org/doi/10.1103/PhysRevE.62.940},
   year = {2000},
   type = {Journal Article}
}

@article{Geoghegan1995,
   author = {Geoghegan, M. and Jones, R. A. L. and Clough, A. S.},
   title = {Surface directed spinodal decomposition in a partially miscible polymer blend},
   journal = {The Journal of Chemical Physics},
   volume = {103},
   number = {7},
   pages = {2719–2724},
   ISSN = {0021-9606},
   DOI = {10.1063/1.470506},
   url = {https://doi.org/10.1063/1.470506},
   year = {1995},
   type = {Journal Article}
}

@article{Cras1999,
   author = {Cras, J. J. and Rowe-Taitt, C. A. and Nivens, D. A. and Ligler, F. S.},
   title = {Comparison of chemical cleaning methods of glass in preparation for silanization},
   journal = {Biosensors and Bioelectronics},
   volume = {14},
   number = {8},
   pages = {683–688},
   ISSN = {0956-5663},
   DOI = {https://doi.org/10.1016/S0956-5663(99)00043-3},
   url = {https://www.sciencedirect.com/science/article/pii/S0956566399000433},
   year = {1999},
   type = {Journal Article}
}

@article{Arkles1977,
   author = {Arkles, Barry},
   title = {Tailoring surfaces with silanes},
   journal = {Chemtech},
   volume = {7},
   pages = {766–778},
   year = {1977},
   type = {Journal Article}
}

@article{Binder1998,
   author = {Binder, Kurt},
   title = {Spinodal decomposition in confined geometry},
   journal = {Journal of Non Equilibrium Thermodynamics},
   volume = {23},
   number = {1},
   pages = {1–44},
   ISSN = {0340-0204},
   year = {1998},
   type = {Journal Article}
}

@article{Puri1997,
   author = {Sanjay, Puri and Harry, L. Frisch},
   title = {Surface-directed spinodal decomposition: modelling and numerical simulations},
   journal = {Journal of Physics: Condensed Matter},
   volume = {9},
   number = {10},
   pages = {2109},
   ISSN = {0953-8984},
   DOI = {10.1088/0953-8984/9/10/003},
   url = {https://doi.org/10.1088/0953-8984/9/10/003},
   year = {1997},
   type = {Journal Article}
}

@article{Steenhoff2025,
   author = {Steenhoff, Jesse M. and Haase, Martin F.},
   title = {Analysis of bijel formation dynamics during solvent transfer-induced phase separation using phase-field simulations},
   journal = {Physical Chemistry Chemical Physics},
   volume = {27},
   number = {10},
   pages = {5117–5130},
   ISSN = {1463-9076},
   DOI = {10.1039/D4CP04638B},
   url = {http://dx.doi.org/10.1039/D4CP04638B
https://pubs.rsc.org/en/content/articlepdf/2025/cp/d4cp04638b},
   year = {2025},
   type = {Journal Article}
}

@article{Nestler2022,
   author = {Farzaneh Kalourazi, Saeideh and Wang, Fei and Zhang, Haodong and Selzer, Michael and Nestler, Britta},
   title = {Phase-field simulation for the formation of porous microstructures due to phase separation in polymer solutions on substrates with different wettabilities},
   journal = {Journal of Physics: Condensed Matter},
   volume = {34},
   number = {44},
   pages = {444003},
   ISSN = {0953-8984},
   DOI = {10.1088/1361-648X/ac8b4d},
   url = {https://doi.org/10.1088/1361-648X/ac8b4d},
   year = {2022},
   type = {Journal Article}
}

@article{Wang2021,
   author = {Wang, Fei and Nestler, Britta},
   title = {Wetting transition and phase separation on flat substrates and in porous structures},
   journal = {The Journal of Chemical Physics},
   volume = {154},
   number = {9},
   ISSN = {0021-9606},
   DOI = {10.1063/5.0044914},
   url = {https://doi.org/10.1063/5.0044914},
   year = {2021},
   type = {Journal Article}
}

\end{document}